\documentclass[pra,twocolumn,superscriptaddress,longbibliography,floatfix,aps,10pt]{revtex4-2}

\usepackage{anyfontsize}
\usepackage[SquareTraceBrackets]{quantum}
\usepackage{graphicx,bm,natbib,upgreek,amsmath,mathrsfs,accents}
\usepackage{amsbsy}
\usepackage[dvipsnames]{xcolor}
\definecolor{myblue}{named}{MidnightBlue}
\definecolor{mygreen}{RGB}{0,120,0}

\usepackage[T1]{fontenc}
\usepackage{newtxtext,newtxmath}

\usepackage[scaled]{helvet}
\usepackage{tikz}
\usetikzlibrary{arrows,decorations.pathmorphing,backgrounds,positioning,fit,petri}
\usepackage{gensymb}				
\usepackage[caption=false]{subfig}
\usepackage{enumitem}
\usepackage{tabularx}
\usepackage{multirow}
\usepackage{bigdelim}
\usepackage[english]{isodate,babel}
\usepackage{optidef}
\usepackage{times}

\DeclarePairedDelimiter\floor{\lfloor}{\rfloor}

\makeatletter
\def\thickhline{%
  \noalign{\ifnum0=`}\fi\hrule \@height \thickarrayrulewidth \futurelet
   \reserved@a\@xthickhline}
\def\@xthickhline{\ifx\reserved@a\thickhline
               \vskip\doublerulesep
               \vskip-\thickarrayrulewidth
             \fi
      \ifnum0=`{\fi}}
\makeatother

\usepackage{amsthm}

\DeclareGraphicsExtensions{.pdf, .jpg, .eps, .svg}

\newcommand{\je}[1]{\textcolor{black}{#1}}

\usepackage[colorlinks=true,linkcolor=teal,citecolor=teal,urlcolor=teal]{hyperref}

\begin{document}

\title{Operational performance of finite-key entanglement-based satellite quantum key distribution}

\author{Jasminder S. Sidhu}
\email{jsmdrsidhu@gmail.com}
\thanks{Corresponding author email}
\affiliation{SUPA Department of Physics, University of Strathclyde, Glasgow, G4 0NG, United Kingdom}
\affiliation{School of Mathematical and Physical Sciences, University of Sheffield, Sheffield, S3 7RH, United Kingdom}
\affiliation{Dahlem Center for Complex Quantum Systems, Freie Universit{\"a}t Berlin, 14195 Berlin, Germany}
\author{Sarah E. McCarthy}
\affiliation{Fraunhofer Centre for Applied Photonics, Glasgow, G1 1RD, United Kingdom}
\affiliation{SUPA Department of Physics, University of Strathclyde, Glasgow, G4 0NG, United Kingdom}
\author{Cameron Paterson}
\affiliation{SUPA Department of Physics, University of Strathclyde, Glasgow, G4 0NG, United Kingdom}
\author{Nathan Walk}
\affiliation{Dahlem Center for Complex Quantum Systems, Freie Universit{\"a}t Berlin, 14195 Berlin, Germany}
\author{Jens Eisert}
\affiliation{Dahlem Center for Complex Quantum Systems, Freie Universit{\"a}t Berlin, 14195 Berlin, Germany}
\author{Daniel K. L. Oi}
\affiliation{SUPA Department of Physics, University of Strathclyde, Glasgow, G4 0NG, United Kingdom}

\date{\today}

\begin{abstract}
Establishing information-theoretically secure communication between arbitrary locations on Earth is a central goal of quantum networking, for which space-based distribution of quantum entanglement is a leading approach. Modelling and analysis of the performance of satellite entanglement pair distribution is therefore important for the architecture and design of future constellations and space systems. However, entanglement-based quantum key distribution in the absence of quantum repeaters, is especially prone to finite-key effects due to low coincident count rates compared to trusted-node single-path links. Quantifying these effects under realistic orbital conditions is therefore essential for the design of practical satellite quantum communication systems. Here, we develop a high-fidelity model of pair distribution from a low-Earth-orbit satellite that captures orbital dynamics, elevation-dependent loss, background noise, and extraneous detector effects. Integrating this with a finite-key security framework for the BBM92 protocol, we optimise the secret key length across different overpass geometries, orbital altitudes, and optical ground station (OGS) separations. We show that secure key generation remains feasible in early clear daytime conditions using baseline parameters representative of current technologies, while showing that modest system improvements could further extend daytime operation. Finally, we evaluate the annual secret key performance for representative intercity-scale OGS pairings. Our results provide quantitative performance bounds and practical design guidelines for near-term entanglement-based satellite QKD, enabling informed trade-offs between satellite payload complexity, ground infrastructure, and achievable secure key throughput.
\end{abstract}

\maketitle

\section{Introduction}
\label{sec:intro}

\noindent
\je{Quantum communication aims to enable information-theoretically secure communication between arbitrary locations on Earth, a long-standing benchmark for the field and a key milestone towards a global quantum internet~\cite{sidhu2021advances, belenchia2021quantum}. More broadly, quantum networking interconnects quantum processors, sensors, and communication nodes into scalable architectures~\cite{sidhu2021advances, belenchia2021quantum}, allowing entanglement to be distributed as a fundamental resource for communication, computation, and sensing~\cite{Pant2019_NPJQI, Lee2022_npjQI, Gottesman2012_PRL, Sidhu2020_AVS, Malia2022_N, Liu2024_NC, Main2025_N}. Realising this vision requires not only reliable entanglement distribution, but also robust link availability, network synchronisation, and coordinated operation across the quantum network stack~\cite{Bacciottini2025_IEEE}. A major obstacle to global-scale quantum communication is the exponential attenuation of photons in optical fibres, which limits direct transmission to distances of approximately $100\,\mathrm{km}$~\cite{Pirandola2017_NC}. While approaches such as twin-field quantum key distribution extend this range~\cite{Lucamarini_2018}, quantum repeaters are widely regarded as the most promising route towards overcoming the fundamental \emph{Pirandola-Laurenza-Ottaviani-Banchi} (PLOB) bound for lossy quantum communication channels and enabling long-distance quantum networking~\cite{Pirandola2017_NC,PhysRevResearch.4.023158}.}

Free-space quantum optical links can \je{effectively} circumvent exponential fibre losses as well as provide more flexible connectivity~\cite{airqkd2022}. In particular, space offers a vantage point for extending quantum links to intercontinental scales through satellite-based communication~\cite{QKDreview2020, Lu2022_RMP, McArthur2025_SPIE, Goswami2025_OQ}. Satellite QKD (SatQKD) has been well studied in a trusted-node configuration, in which the satellite is regarded as a trusted station that can carry out separate QKD protocols with various ground stations. Although this approach is attractive in terms of attractive in terms of practical satellite payloads, it does require us to abandon the traditional level of security delivered by QKD, whereby the everything outside of the stations directly controlled by Alice and Bob is regarded as an potentially hostile channel.

\je{Entanglement comes to the rescue: by distributing non-local quantum correlations across global distances, \emph{entanglement-based satellite quantum key distribution} (Ent-SatQKD) provides one of the most promising routes towards secure worldwide quantum communication.} Ent-SatQKD
serves both as a near-term demonstrator and a quantitative benchmark for quantum networking, enabling end-to-end validation of entanglement distribution and free-space channel performance for secure key generation under realistic conditions~\cite{Bourgoin:2013fk, sidhu2021advances}. Entanglement-based protocols, such as the 
\je{\emph{Bennett-Brassard-Mermin92}
(BBM92) scheme}, derive their security solely from non-local quantum correlations, enabling secure communication in \je{the absence of trusted intermediate nodes}~\cite{Bennett1992_PRL}. 

\je{Although major milestones have already been achieved, realizing a global quantum communication network still requires substantial advances in both methodology and technology.}
The Micius satellite has already established the feasibility of distributing entanglement and performing \emph{quantum key distribution} (QKD) over 1,000~km~\cite{jianwei2018progress, Yin2017_S}, with several other dedicated satellite missions following~\cite{Villar2020, islam2024finite, Haber2018_Qubeproceedings,sivasankaran2022, Jennewein2014_SPIE,Hiemstra_2025,Zhang24,rao2025quantumrandomnumbergeneration}. In parallel, advances in entangled-photon sources and quantum memories have established the theoretical groundwork for space-based quantum repeater architectures, outlining the hardware and protocol layers required for global entanglement distribution~\cite{Boone2014, Liorni2021, Gundogan2021_NPJQI, gundogan2021topical, Wallnofer2022, Gundogan2024_OQ, paterson2026singlesatellitequantumrepeaterperformance}. Complementing these developments, hybrid protocols that combine discrete- and continuous-variable approaches are being developed to enhance the integration of space- and terrestrial-based quantum networks~\cite{Qi2021_PRA,Primaatmaja2022_Q,Sidhu2025_IoP}. \je{Taken together, these developments have transformed global quantum networking from a long-term vision into a concrete technological roadmap. Nevertheless, several key challenges remain}

In contrast to fibre links, satellites in \emph{low-Earth orbit} (LEO) have limited contact times with an \emph{optical ground station} (OGS) and operate with 
severe \emph{size, weight, and power} (SWaP) constraints~\cite{Sidhu2023finite}. These constraints place Ent-SatQKD in a 
\je{delicate} finite-resource regime, where statistical fluctuations in channel estimates dominate key generation and restrict the operational margins of small-satellite missions~\cite{islam2024finite}. While trusted-node SatQKD has both been analysed in the finite-resource framework~\cite{sidhu2021key,Sidhu2022_npjQI,Roger:2023,Crampton:2026} and made significant strides towards implementation~\cite{Liao2017_N,Yin2017_S,Kerstel2018_EPJ, Mazzarella2020_C,Villar2020,Yin2020_N}, untrusted-node Ent-SatQKD remains comparatively less explored despite its direct relevance for scalable entanglement-based quantum networks~\cite{Lim2021_PRL, Panigrahy2022_IEEE, Zaunders2025_arxiv,Bacciottini2025_IEEE, koudia2024spacebasedquantuminternetentanglement} that can realise the full security guarantees of standard QKD. \je{Understanding its performance under realistic orbital dynamics and finite-resource constraints is therefore a central challenge on the path towards practical global quantum networking and the distributed quantum technologies it promises}~\cite{Zhang_2021, Main_2025,huang2026quantum}.

We address this gap by developing a comprehensive end-to-end model of dual-downlink BBM92 SatQKD that integrates orbital dynamics, elevation-dependent losses, and realistic detector behaviour. Coupled to a rigorous finite-key security framework, this model enables quantitative assessment of achievable secure key lengths for untrusted-node operation. We apply the framework to target a number of near- and long-term questions. First, we examine how practical limitations and operational conditions, such as daylight operation, impacts achievable key yields. Then, we establish performance bounds for dual-downlink BBM92 SatQKD across a representative range of OGS separations and overpass geometries compatible with existing technologies. Third, by analysing the long-term averaged key generation, we determine trade-offs between performance and hardware overheads. Ultimately, this allows us to carry out an extensive, multi-parameter optimisation to identify the best overpass strategies revealing substantial differences between the dual-downlink considered here and single downlink configurations. This work advances the capability to design and optimise future space-based quantum communication networks. The BBM92 finite-key rates derived here serve as a benchmark for secure key generation and as quantitative proxies for entanglement distribution capacity for global-scale quantum networking applications.

The remainder of this paper is structured as follows. Section~\ref{sec:background} introduces our system model and finite-key optimisation, outlining the full orbital geometry, elevation-dependent losses, and detector noise mechanisms that define the dual downlink BBM92 configuration and the secret key lengths. Section~\ref{sec:results} outlines the optimised secret keys under different operational scenarios: specifically in section~\ref{subsec:block_size} we illustrate the importance of optimising key lengths using only specific segments of overpass transmission data. Section~\ref{subsec:SKL_altitude_d_variation} outlines finite-resource performance for different overpass geometries, while in section~\ref{subsec:annual_SKL} we determine the annual performance of a SatQKD system. Conclusions and discussions are provided in section~\ref{sec:conc}, where we discusses the implications of these findings for near-term Ent-SatQKD missions and for the scalable deployment of space-based entanglement distribution networks.


\section{Background and system model}
\label{sec:background}

\noindent
We model the BBM92 secret key length for a dual downlink QKD channel operating at night, where the satellite is in a circular \emph{low Earth orbit} (LEO) of altitude 500~km. The three main elements are: (i) a system model that captures full dynamics of a satellite overpass geometry in relation to two OGSs, (2) an estimate of the expected click and loss statistics, or data blocks received at both OGSs, and (3) a rigorous account of finite block sizes that reliably returns a secure key length from the data blocks. We outline (1) and (2) in detail in section~\ref{sec:system_model}, and (3) in section~\ref{subsec:protocol}. Section~\ref{subsec:optimisation} outlines our approach to optimising key rates. Our high-fidelity model can estimate the sensitivity of key rates against a range of experimentally informed parameters to assess the feasibility of near-term implementation. Our model also permits a generalisation to model a range of different system models, providing a timely utility to explore trade-offs in competing mission designs.


\subsection{System model}
\label{sec:system_model}

\begin{figure}
    \centering
    \includegraphics[width=\columnwidth]{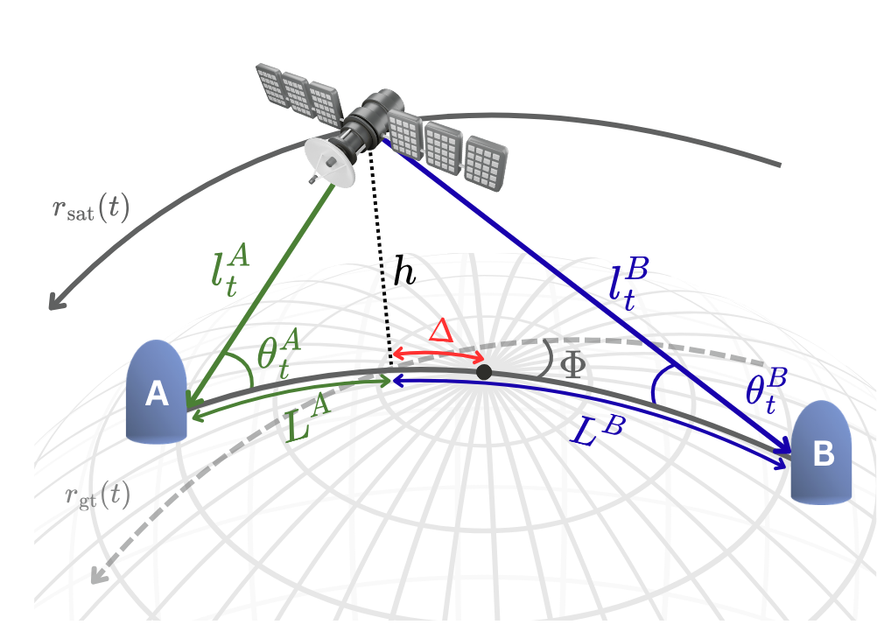}
    \caption{\textbf{Dual downlink satellite overpass geometry.} Depiction of a satellite geometry for an asymmetrical overpass at the $t_0$ point. The orbit can be described through the relation between the satellite's ground track and the OGS baseline, or more specifically the angular rotation $\Phi$ and the distance offset, $\Delta$ from the baseline midpoint. The OGSs shown are located equidistant from the North Pole marked by a black dot. $L^A$ and $L^B$ are the distances between the ground stations such that the baseline distance $d=L^A+L^B$.}
    \label{fig:geom}
\end{figure}

\noindent
We begin by introducing our system model. Consider the simultaneous distribution of entangled photon pairs from a satellite in a circular orbit (altitude $h$) and two \emph{optical ground stations} (OGS), $\text{OGS}_A$ and $\text{OGS}_B$, separated by distance $d$ along a great-circle arc segment (the OGS baseline). The overpass geometry can be characterised by two parameters (Fig.~\ref{fig:geom}). The first, $\Phi$, is the intersection angle between the orbital ground track and the OGS baseline. The second, $\Delta$, is the distance between the midpoint of the OGS baseline and the ground track intersection point, characterising the asymmetry of the overpass.  We denote $L_{A/B}(t)$ as the distances between the intersection point of the ground track with the OGS baseline intersection and each OGS.

From this parameterisation, we identify four distinct overpass geometries. First, when $\Phi = 0^\circ,\ \Delta = 0$~km (zenith, symmetric), we have an ideal zenith overpass where the satellite passes directly over each OGS consecutively. In contrast to the single OGS scenario (such as for prepare-and-measure BB84 protocols~\cite{Sidhu2022_npjQI}), the total channel loss may display 2 minima during the overpass, due to the dual path geometry and the inverse square law. For $h=500$~km this effect arise for OGS separations $>950$~km. Second, when $\Phi = 90^\circ$ and $\Delta = 0$~km (non-zenith, symmetric), the satellite passes directly between the ground stations with $L_A(t) = L_B(t)$. Third, when $\Delta = \pi R/2$, where $R$ is the Earth's radius, the ground track is symmetrically offset from OGS baseline. Finally, when both $\Phi\neq0^\circ$ and $\Delta\neq0$ (non-zenith, asymmetric), we have the most general overpass geometry.

With this parameterisation, we determine the instantaneous slant range $l^{i}_t$ and elevation angles $\theta^i_t$ between the satellite and OGS$_i$ as a function of time (Methods~\ref{subsec:link_geom}). We impose a minimum elevation of $\theta_\text{min} = 10^\circ$ for both links to account for possible visibility. Within the duration of simultaneous visibility, the instantaneous loss for each link $i = \{A, B\}$ is
\begin{align}
    \eta_\lambda^i(t) = \eta_\text{diff}^i(\lambda, \theta^i_t) \, \eta_\text{atm}^i(\lambda, \theta^i_t) \, \eta_\text{int}\, ,
\end{align}
where $\eta_\text{diff}$, $\eta_\text{atm}$, and $\eta_\text{int}$ are losses from diffraction, atmospheric scattering and absorption, and a fixed `intrinsic' system efficiency respectively (Methods~\ref{subsec:loss_modelling}). Losses due to diffraction effects use the Fraunhofer approximation to the Rayleigh-Sommerfeld diffraction integral to estimate the ratio of the power transmitted from the satellite to that received at the ground station. MODTRAN~\cite{Modtran_inproceedings} provides the elevation-dependent link efficiency due to atmospheric effects for the location of the ground station and the wavelength of the photon pairs. The intrinsic loss is taken to be the same for both links, such that the combined loss is simply $\eta_t = \eta_\lambda^A(t)\eta_{\lambda'}^B(t)$. For the entanglement distribution, we can consider cases where the photon pairs are wavelength degenerate or non-degenerate.

\begin{figure}[t!]
    \centering
    \includegraphics[width=\columnwidth]{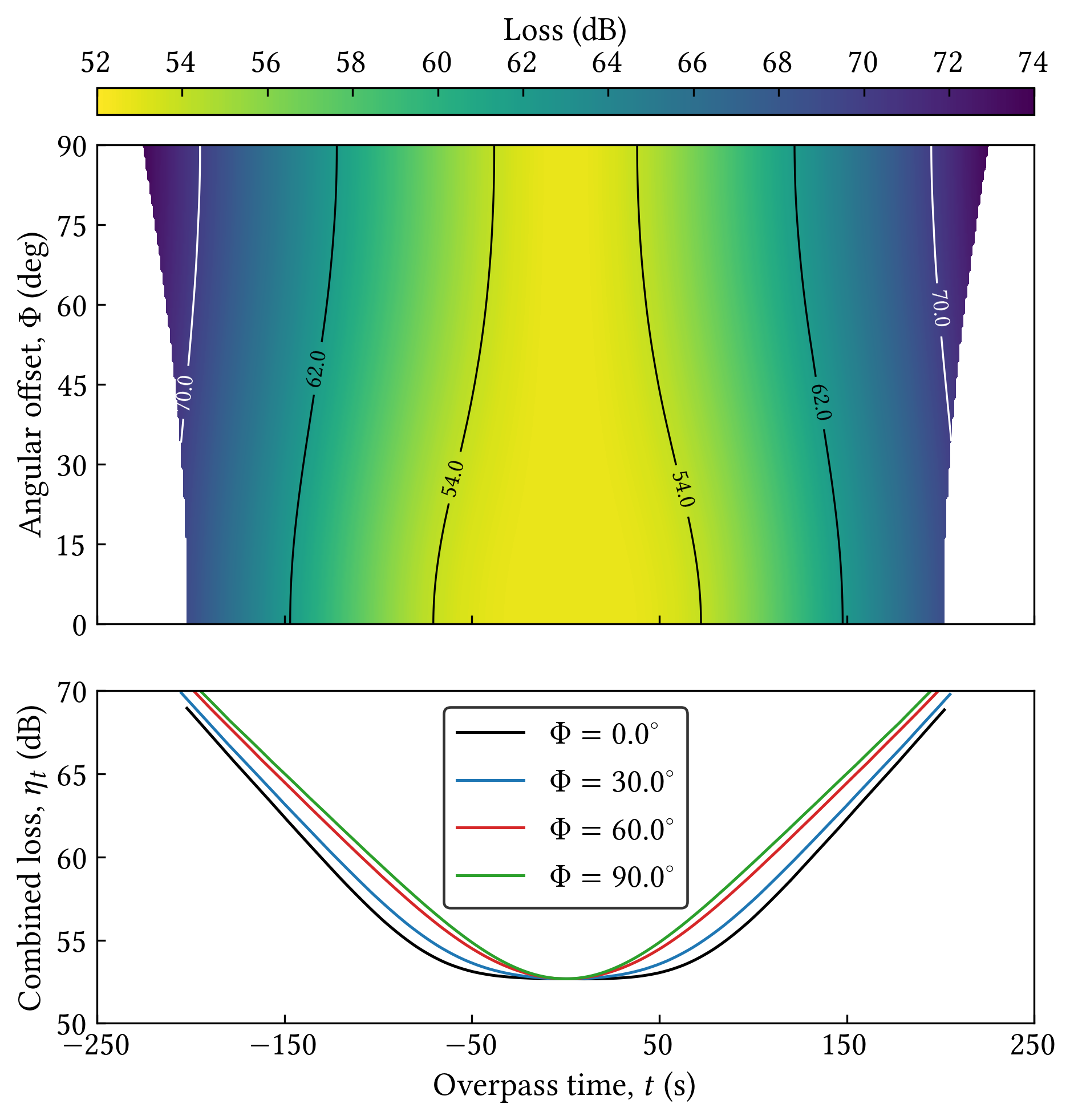}
    \caption{\textbf{Link model for satellite-to-ground QKD}. (top) Instantaneous link efficiency for satellite overpasses within the visible region with $\Delta=0$~km and varying $\Phi$, with $\lambda=785$ nm and OGS separation of $d = 1000$~km. The white region illustrates where either satellite link elevation falls below $\theta_\text{min}=10^\circ$. Note that the smallest transmission time corresponds to the satellite passing directly overhead both OGSs, where the ground track traces the great circle arc between the OGSs. Further, the two loss minima are due to the inverse square law scaling of the dual downlinks and occurs at large $d$. (bottom) Combined loss vs overpass time for different $\Phi$, illustrating a minimum loss of 53~dB for a zenith overpass ($\theta^i_\text{max} = 90^\circ$) for $i \in \{A, B\}$. All other system parameters in Table~\ref{tab:system_parameters}.}
    \label{fig:loss}
\end{figure}%

From the link loss, we determine the click statistics from the simultaneous detection of photons at both OGSs. We assume a continuous wave pumped spontaneous parametric down conversion source of polarisation-entangled photons with a 100~Mpair~s$^{-1}$ pair rate and an intrinsic QBER of 0.5\%, corresponding to e.g. imperfect Bell state fidelity. We also account for errors owing to incorrect state discrimination in the receiver optics, extraneous counts from dark counts, afterpulsing, and background light detected by the ground station telescope. The distinct combination of joint clicks from signal and extraneous count are also included. Since background light is site-specific, we determine the probability of background clicks separately each for OGS using radiance data from MODTRAN and assuming nighttime operation together with the nominal system characteristics summarised in Table~\ref{tab:system_parameters}. Once the click statistics are estimated, we account for a factor of $1/2$ to establish the post sifting block size. Full details on data and error block statistics are provided in Methods~\ref{subsec:ext_counts} and~\ref{subsec:block}.

\newcommand*{\tabindent}{ \hspace{-1mm}}
\newlength{\thickarrayrulewidth}
\setlength{\thickarrayrulewidth}{2.1\arrayrulewidth}
\renewcommand{\arraystretch}{1.25}
\setlength{\tabcolsep}{8pt}
\begin{table}[t!]
  \centering
  \begin{tabular}{m{3.9cm}|m{1.3cm}|m{1.87cm}}
    \thickhline
    \textbf{Parameter description} 				& \textbf{Notation}			& \textbf{Value} 	\\
    \hline
    OGS separation                              			&   	$d$              			& 1000 km        		\\
    Transmitter aperture diameter  				&	$T_\mathsf{X}$			& 10 cm			\\
    Receiver aperture diameter					&	$R_\mathsf{X}$			& 70 cm			\\
    Gaussian Beam waist						&	$w_0$				& 5 cm			\\
    Source wavelength						&	$\lambda$				& 785 nm			\\    
    Source rate		 						&	$f_s$				& $100$~Mpair~s$^{-1}$		\\ 
    Satellite orbit altitude	 					&	$h$					& $500$ km 		\\ 
    Minimum elevation limit						&	$\theta_\text{min}$		& $10^\circ$		\\
    Intrinsic quantum bit error rate				& 	$\text{QBER}_\text{I}$	& 0.5\% 			\\ 
    Dark count probability per bin				&	$p_\text{dc}$			& $5\times 10^{-7}$ 	\\ 
    After-pulsing probability	  					&	$p_\text{ap}$			& 0.1\% 			\\ 
    Spectral filter bandwidth    					&      $\Delta\lambda$ 		& 10~nm 			\\
    Full angular field of view 					& 	FOV & $10$~$\mu$rad\\ 
    Coincidence window 						& 	$\Delta\tau$ 			& 500~ps 			\\
    Single channel loss at zenith					&	$\eta^{i}_\text{loss}$		& $22.70$ dB 		\\ 
    \textcolor{black!60!white}{
    $\hookrightarrow$Diffraction loss at zenith}		&	\textcolor{black!60!white}{$\eta_\text{diff}^i(\lambda, 90)$}	& \textcolor{black!60!white}{8.47~dB} \\
    \textcolor{black!60!white}{$\hookrightarrow$Atmospheric loss at zenith}	&	\textcolor{black!55!white}{$\eta_\text{atm}^i(\lambda, 90)$}	& \textcolor{black!60!white}{0.73~dB} \\
    \textcolor{black!60!white}{$\hookrightarrow$Optical inefficiency}		&	 \multirow{2}{*}{\textcolor{black!60!white}{$\bigg\}$ $\eta_\text{int}$}}				& \multirow{2}{*}{\textcolor{black!60!white}{$13.5$~dB}}		\\ 
    \textcolor{black!60!white}{$\hookrightarrow$Imperfect beam propagation}	&	& 				\\ 
    \color{black}
    Security level						    	&	$s$					& 6		 		\\     
    Correctness parameter 						&	$t$		             		& $\log_2 10^{s+2}$ 	\\ 
    \thickhline
  \end{tabular}
  \caption{\textbf{Reference system parameters}. Transmitter, receiver, and source properties determine range and elevation-dependent loss. The single channel loss is given for the as the zenith occurs at a different position for $i=A,B$, however the value will be the same. This is broken down into the components from diffraction and atmospheric absorption. The `intrinsic' system loss is broken down into two components (Methods~\ref{subsec:loss_modelling}). The combined $\eta^\text{tot}_\text{loss}$ can be scaled to model other SatQKD systems that differ by a fixed link loss ratio, e.g. different $T_\mathsf{X}$ or $R_\mathsf{X}$ apertures, or detector efficiencies. The intrinsic quantum bit error rate, $\text{QBER}_\text{I}$, incorporates errors from source quality, receiver measurement fidelity, basis misalignment, and polarisation fluctuations, while the extraneous count probability, $p_\text{ec}$, incorporates detector dark count and background rate. The correctness and security parameters are used to determine the finite-block composable SKL.}
  \label{tab:system_parameters}
\end{table}%
%


\subsection{The protocol and secret key length}
\label{subsec:protocol}

\noindent
The BBM92 protocol considers an entangled state shared between two parties that is derived from a larger system potentially entangled with an adversary. Each party performs ideal, incompatible qubit measurements in randomly chosen bases. After the measurement stage, the parties publicly announce their basis choices over a classical channel and sift out rounds where different bases were chosen. The sifted data is then divided into two subsets: one for parameter estimation and one for raw key generation. The protocol deviates from the original BBM92 formulation by implementing random sampling without replacement to allocate bits between these tasks. This symmetrisation ensures the uniform distribution of errors across both subsets, enabling the use of finite-size statistical bounds to estimate the error rate of the raw key from that of the sample. A protocol run only proceeds if the sample error rate is below a threshold $\delta$, preserving the security of the remaining data. In the finite-key analysis, the portion of the sifted data block length $m$ dedicated to parameter estimation is denoted by $\beta$, such that the sample size is $k=\floor{\beta m}$. The statistical estimation further depends on a tolerated excess error $\nu$, which quantifies how much larger the raw key error rate could be compared to $\delta$ while still passing the test, and on a refinement parameter $\xi$, which bounds this excess when applying the improved sampling inequality. Together, these optimisation parameters determine the trade-off between using more data for parameter estimation (increasing confidence in the test) and retaining more data for the final secret key.

Denoting the finite key rate $\alpha = \ell /m$, the maximised BBM92 finite-key length over the four-dimensional real vector $\bm{x} = (\alpha, \beta, \nu, \xi)$, is the solution to the following optimisation program~\cite{Lim2021_PRL}
\begin{maxi}|s|
{\bm{x} \in \mathbbm{R}^4}{\ell = \lfloor \alpha m \rfloor \, ,}
{}{}
\addConstraint{2^{-t} + 2 \epsilon_\text{pe}(\nu, \xi) + \epsilon_\text{pa}(\nu)}{\leq \epsilon_\text{QKD} \,}
\addConstraint{0 \leq \alpha \leq 1\, , \quad 0 < \beta}{\leq \frac{1}{2} \,}
\addConstraint{0 < \xi < \nu}{< \frac{1}{2} - \delta \, ,}
\label{opt_finite_key}
\end{maxi}
The first non-trivial constraint determines the maximum achievable secret key length that is $\epsilon_\text{QKD} = 10^{-s}$ secure, subject to some tolerated excess error rate, $\nu$, between the observed error rate in the parameter estimation sample and true error rate in the raw key~\cite{Tomamichel2017_Q}. The error functions due to privacy amplification and parameter estimation that describe the protocol security are defined  
\begin{align}
\begin{split}
\epsilon_\text{pe}(\nu, \xi)^2 &= \exp\hspace{-0.2em}\left[-\frac{2mk\xi^2}{n+1}\right] +\exp\hspace{-0.2em}\left[-2\Gamma_{m(\delta+\xi)}[(n\nu')^2 - 1]\right] \\
\epsilon_\text{pa}(\nu) &= \frac{1}{2} \sqrt{2^{-n (1-h_2[\delta+\nu]) + r + t + \ell}}\, .
\end{split}
\end{align}
respectively, with $k=\floor{\beta m}$, $\Gamma_{x,m}=1/(x+1) + 1/(m-x+1)$, $\nu' = \nu-\xi$, $t=\log_2 10^{s+2}$ the correctness error, $r = 1.19 n h_2(\delta)$ the estimated error correction leakage, and $h_2(x)=-x\log_2 x -(1-x)\log_2(1-x)$ is the binary entropy function. Intuitively, the excess error rate $\nu$ provides a
safety margin to guard against statistical fluctuations in finite-size data, while $\xi$ partitions these fluctuations into two regimes to enable a tighter bound on the probability that the parameter-estimation test 
passes despite the raw key exceeding the tolerated error rate. The quantity $\epsilon_\text{pa}(\nu)$ is determined by the leftover hash lemma, where the smoothing 
term in the min-entropy bound depends on $\nu$ through the estimated phase error rate, while the quantity $\epsilon_\text{pe}(\nu,\xi)$ is determined by the probability that the parameter-estimation test passes even though the raw key error rate exceeds $\delta+\nu$. Tuning the parameters $\bm{x}$ over this feasible region defined through the remaining constraints in Eq.~\eqref{opt_finite_key} optimises the trade-off between key length and statistical confidence that maximises the key length for a given data block and system parameters.


\subsection{Secret key rate optimisation}
\label{subsec:optimisation}

\noindent
Given the complex, \je{non-convex nature of the parameter space, objective function and complex inequality constraints}, we adopt a brute force search strategy to solve the optimisation in Eq.~\eqref{opt_finite_key}. This guarantees reliable coverage that does not depend on objective function smoothness or gradients compared with direct optimisation methods. To improve the search efficiency, we analytically reformulate the optimisation problem to remove one optimisation parameter. Specifically, from the non-trivial security constraint in Eq.~\eqref{opt_finite_key}, we isolate the upper bound on key length
\begin{align}
\log_2 \left[4(10^{-s} - 2^{-t} - 2 \epsilon_\text{pe}(\nu,\xi))^2\right] + n (1 - h_2(\delta + \nu)) - r - t\, ,
\label{eq:opt_finite_key_2}
\end{align}
which reduces the parameterisation by eliminating $\alpha$ from the search space. For a given set $\bm{x}_\text{opt} = \{\beta, \nu, \xi\}$ this upper bound cannot be exceeded without violating security. If the constraint holds with slack, this bound may be larger than the secret key length produced by the protocol, but in the optimal setting the slack is zero and the upper bound tight. The trivial bounds in Eq.~\eqref{opt_finite_key} are enforced by restricting the search space so that they are satisfied by construction, while the non-trivial constraint is still checked for each parameter triplet $\bm{x}_\text{opt}$. For all admissible points, the corresponding key length is computed, and the maximum secret key length and its associated triplet is stored as the final optimisation result.

By eliminating $\alpha$, we reduce the dimensionality of the parameter domain to provide a computation speedup. Assuming that each optimisation parameter is sampled $n$ times, the search complexity is reduced from $\mathcal{O}(n^4)$ to $\mathcal{O}(n^3)$. Furthermore, parameter domain constraints are enforced with efficient parallelisation to achieve a significant computational speedup. Full details of our key rate optimisation routine are provided in the pseudocode in Fig.~\ref{fig:protocol_algo}.


\section{Results}
\label{sec:results}

\noindent
To demonstrate the utility of our optimisation framework, we present a comprehensive performance analysis of the finite-key BBM92 protocol under realistic operational conditions. First, for the baseline system configuration outlined in Table~\ref{tab:system_parameters}, section~\ref{subsec:block_size} characterises the intrinsic finite key performance under varying block sizes and background illumination during a single satellite overpass. We motivate the thresholding approach adopted throughout the remainder of the analysis for overpass block optimisation and assess the protocol’s robustness to background noise. 

Second, section~\ref{subsec:SKL_altitude_d_variation} investigates the impact of network geometry on key generation by varying the OGS separation and satellite altitude relative to the nominal baseline scenario. This analysis quantitatively addresses the number of OGS sites required to provide key service within the satellite’s line of sight, thereby informing hardware overheads and the tradeoff between satellite altitude and OGS separation. 

Third, section~\ref{subsec:annual_SKL} extends the analysis beyond single satellite overpasses to evaluate performance across varying overpass geometries and estimate the annual secret key length, providing a system-level assessment of long-term performance. Together, these results quantify both the instantaneous and aggregate capabilities of the protocol in realistic entanglement-based quantum communication scenarios. We conclude the results section with a discussion of enabling technologies in section~\ref{subsec:space_comp_ent_sources} required to deliver the modelled performance.


\subsection{Protocol performance and thresholding}
\label{subsec:block_size}

\noindent
With SatQKD having limited operating margins in the finite key regime, we first examine the optimised key length, Eq.~\eqref{eq:opt_finite_key_2}, as a function of the block size. The block size depends on the fraction of the satellite overpass used for key generation. Previous studies typically control the block size by selecting a time window, $\Delta t$, around the satellite's maximum elevation to construct $m(\Delta t)$. This approach is well suited to a single satellite-OGS system, where the loss profile is symmetric about the overpass midpoint. For the BBM92 protocol, however, the loss profile associated with two spatially separated OGSs is generally asymmetric. Selecting data around the midpoint is therefore generally suboptimal, since the overpass QBER minima may occur in a different portion of the overpass. Intuitively, an alternative effective block construction scheme should favour data with lower QBERs over a symmetric temporal segment.

To address this, we construct block sizes by imposing a threshold, $\delta$, on the QBER of the data used for key generation. From the estimated time-dependent coincidence counts $D_t$ and error blocks $E_t$ (Methods~\ref{subsec:block}), we compute the instantaneous weighted block QBER, $\smash{\sum_t E_t / \sum_{t'} D_{t'}}$, which defines the minimum and maximum achievable block QBERs over the overpass. By uniformly sampling threshold values within the permissible interval, $\smash{[\text{QBER}_\text{min}, \text{QBER}_\text{max}]}$, we extract corresponding click statistics, $m(\delta)$. Increasing $\delta$ admits a larger fraction of the overpass data, thereby increasing the block size at the cost of a higher average QBER. This threshold-based approach is independent of any symmetry in the overpass loss profile and is therefore more general than time-window segmentation. Further details and intuition are provided in Methods~\ref{subsec:thresholding}. 

To assess the finite-key performance of BBM92 and its robustness to background light, we evaluate the SKL as a function of block size under varying background levels. For each link $i \in \{A,B\}$, we define the time-dependent extraneous count probability as
\begin{align}
p_{\text{ec}, t}^i = p_\text{dc} + p_{\text{bg},t}^i - p_\text{dc} p_{\text{bg},t}^i\, , 
\label{eqn:ec_prob}
\end{align}
where $p_\text{dc}$ the dark count probability and $p_{\text{bg},t}^i$ the probability of detecting a background photon at OGS $i$. Per coincidence window, we take representative values of $p_\text{dc}=5\times10^{-7}$ (1000 counts per second) while $p_{\text{bg},t}^i$ is determined from MODTRAN-based radiance simulations (Methods~\ref{subsec:ext_counts}), accounting for site-dependent background variations. To model different illumination conditions, we scale $\smash{p_{\text{bg},t}^i}$ at both OGSs by a common factor $f$. Ref.~\cite{ErLong2005_IOP} serves as a guide on how each background count probability, and hence $f$, maps to different time-of-day conditions. For the baseline parameters in Table~\ref{tab:system_parameters}, a unit factor corresponds to moonless night-time operation, with increasing values representing progressively brighter conditions. In particular, $f=100$ corresponds to full moon night, while values in the range $10^3 \leq f \leq 25 \times 10^3$ to twilight conditions (astronomical, nautical, and civil), where scattered solar photons increasingly dominate the detector background but secure key generation may still be feasible for favourable geometries. Values $f \geq 50\times 10^3$ correspond to early daylight operation.

\begin{figure}
    \centering
    \includegraphics[width=1\columnwidth]{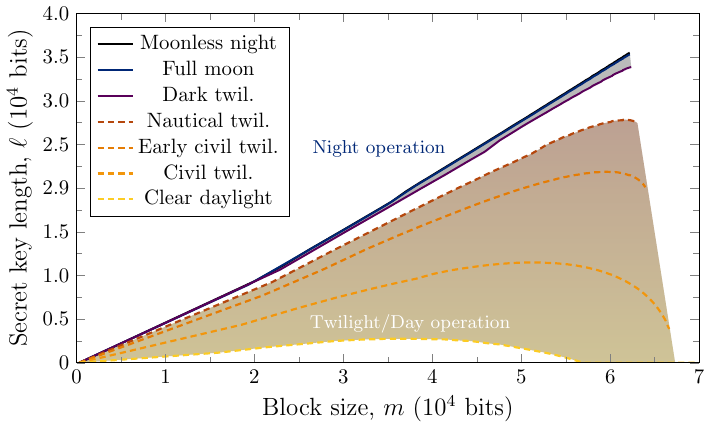}
    \caption{\textbf{Performance with block size and background illumination.} Secret key length, $\ell$, as a function of block size for increasing background photon levels, parameterised by a scaling factor $f$ relative to a baseline background detection probability $\smash{p_\text{bg}=10^{-7}}$. Curves correspond to night-time and twilight (twil.) operating conditions, with darker solid colours indicating lower background levels and progressively lighter dashed colours representing increasing ambient illumination. Shaded regions highlight the night-time, twilight and daylight regimes. Data corresponds to a zenith-zenith overpass, with $h = 500$~km and $d = $1000~km and from top down, the factors are $f=1$, 100, 1000, 5000, 10,000, 25,000, 50,000 respectively.}
    \label{fig:block_size}
\end{figure}

Fig.~\ref{fig:block_size} illustrates the optimised SKL as a function of block size for each background level. In all cases, the SKL initially increases with block size as statistical fluctuations are reduced and parameter estimation becomes more efficient. During night time operation ($f\leq100$, region shaded in black), the SKL remains linear across the full range of block sizes, indicating that the entire overpass data should be used to maximise the key length. Note that for our system parameters, we observe no distinguishable difference between moonless night and full moon performance. In contrast, under twilight ($f=10^3-10^4$), and daylight ($f = 50\times10^3$) conditions, corresponding to the sunrise-coloured region, the SKL exhibits a maximum at an intermediate block size and decreases thereafter. Although admitting more data increases the block size, it does so by also introducing significantly higher-QBER events associated with portions of the overpass where either OGS or both see low elevations. In this regime, maximising the SKL requires joint optimisation over the block size in addition to the protocol parameters, $\text{SKL} = \sup_{\delta} \ell [m(\delta)]$, reflecting a trade-off between data volume and error rate. This behaviour is consistent with earlier finite-key analyses of prepare-and-measure (P\&M) protocols, such as BB84, where small or poorly conditioned data blocks levy a substantial penalty on the achievable key length~\cite{Sidhu2022_npjQI,Sidhu2023finite}. In what follows, we report block-size optimised key lengths.

These results indicate that modest, experimentally feasible adjustments to system parameters can further enhance daytime performance and improve robustness to elevated background counts. Tighter spectral filtering would reduce the impact of background light, however the source is based on the performance of SPDC pair generation, which tends to be broader band. The use of filters narrower than 10~nm may reduce the detection efficiency by blocking the output from the entanglement source as well as those from extraneous sources. Another feasible adjustment would be to reduce the coincidence window to suppress the likelihood of an extraneous count being incorrectly identified as originating from a photon pair source, though this may have the effect of reducing the heralding efficiency of true photon pair coincidences by imposing too narrow a window for arrival. Also, the angular field of view chosen here is relatively large. Reducing this would also reduce the $p_{\text{bg},t}^i$ value, allowing for improved tolerance to higher radiance values. Alternatively, improved single photon detectors with very low $p_\text{dc}$ would allow for greater tolerance to $p_{\text{bg},t}^i$, potentially enabling clear daylight operation.

Comparing the performance of BBM92 with BB84 reveals a second notable feature; the BBM92 protocol exhibits greater resilience to channel loss. In particular, BBM92 can yield a non-zero SKL in regimes where the decoy-state BB84 has zero key yield. This observation rises from how loss features in the security analyses associated with each protocol. In BBM92, key generation follows by optimising over all unknown states $\smash{\rho_{AB}}$ conditioned on joint detection events, with channel losses entering primarily as a multiplicative suppression through the coincidence probability, while privacy amplification depends only on the conditional error rates. Therefore, loss suppresses the achievable rate but does not impose a restrictive constraint on the tolerable loss. Instead, P\&M rates require an accurate estimation of the single-photon yield, which becomes increasingly difficult in the finite-block regime, leading to a constrained region of positive key rates.


%
\begin{figure*}[t!]
    \centering
    \includegraphics[width=1\linewidth]{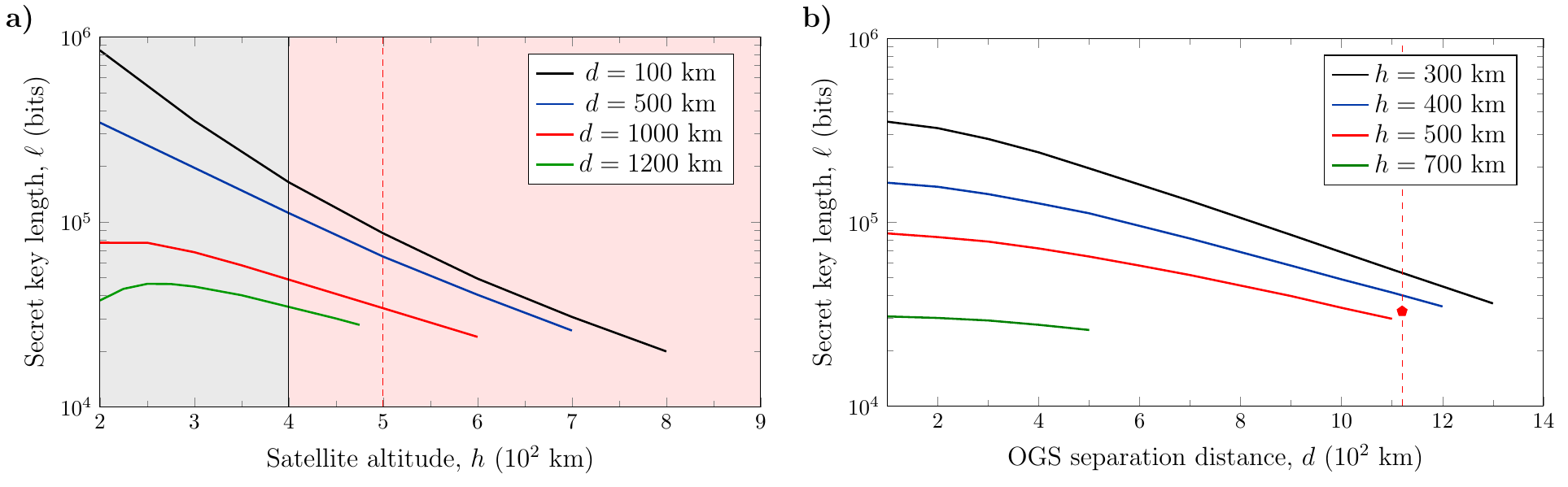}
    \caption{\textbf{SKL with different system configurations}. We consider key generation from a single zenith-zenith overpass $\{\Phi=0^\circ, \Delta=0$~km\}. \textbf{(a)} finite-key BBM92 performance with different satellite altitudes, $h$. The grey region indicates VLEO and the red shaded region LEO orbits. The vertical black line marks the altitude of the international space station and the red dashed line of the Micius satellite. \textbf{(b)} illustrates the key performance with OGS separation distances, $d$. The vertical dashed line indicates the nominal altitude of 500~km with the mark indicating the simulated key length achieved using a subset of parameters aligned to the Micius satellite. Both plots consider a zenith-zenith overpass. System parameters as in Table~\ref{tab:system_parameters}.}
    \label{fig:system_geom}
\end{figure*}%

\subsection{System configuration and overpass geometry analysis}
\label{subsec:SKL_altitude_d_variation}

\noindent
Quantum networking architectures that facilitate entanglement distribution must carefully select the system configuration of the satellite altitude $h$ and the separation distance between OGSs $d$ in order to address trade-offs between entanglement distribution volumes and achievable range. The key distribution performance of each overpass will depend on the geometry of the satellite trajectory with respect to the OGS positions, as described in Section~\ref{sec:system_model}, parameterised by $\{\Phi,\Delta\}$. In this section, we investigate how each parameter influences the secret key length. Fig.~\ref{fig:system_geom}(a) illustrates the secret key length as a function of the satellite altitude for different OGS separation distances for $\{\Phi=0^\circ,\Delta=0$~km\} which shows two main features. First, for a given OGS separation distance, $d$, the secret key length generally decreases with increasing satellite altitude, $h$. This is due to increased channel losses arising from increased diffraction losses and beam broadening, as the link lengths increase with altitude. Only for satellites in a \emph{very low Earth orbit} (VLEO) and widely separated OGSs do we see a local optimum altitude which balances the slant-ranges and simultaneous view times. In this regime, increasing $d$ results in larger slant angles, causing the link to traverse a greater atmospheric path length, the secret key lengths can be improved by increasing the satellite altitude (Fig.~\ref{fig:system_geom}(a) $d=1200$~km, $200-300$~km region). Second, for a given satellite altitude, $h$, we observe the key length also decreases with increasing OGS separation distances, as expected. This trend is similarly driven by increased channel losses, which is dominated by increased atmospheric losses. We note that below the ISS altitude, orbital lifetimes are considerably reduced due to atmospheric drag, hence higher altitudes are preferred for constellation deployment. However, VLEO concepts operating below $200$~km to $300$~km perigee have been proposed based on air-breathing electric propulsion~\cite{andreussi2022review}. With growing commercial interest in VLEO, driven by demand for low-latency communications, and the sustained operation of satellites below 300~km by the China Academy of Space Technology~\cite{Jones2026VLEO}, VLEO is emerging as a viable operational regime. Our results serve as a guide for Ent-SatQKD missions operating in this emerging orbit.

To understand how far two OGSs can be separated to generate non-zero finite keys for a given orbital altitude, Fig.~\ref{fig:system_geom}(b) illustrates the key length as a function of the OGS separation distance, also for $\{\Phi=0^\circ,~\Delta=0$~km\}. The maximum distance between two OGSs within the viewing angle of a satellite at altitude $h$ is $2R\arccos(R/R+h)$~km, ignoring the minimum elevation limit. For a typical LEO orbit altitude of $h=500$~km, we see that keys are generated up to a separation distance of $d =1.1\times10^3$~km, which is approximately 22.5\% of the maximum distance. For similar orbits with altitudes $h=300$~km, $h=400$~km, and $h=700$~km, the percentage of the maximum viewing distance over which keys are generated is approximately 33.9\%, 27.3\%, and 8.7\% respectively. These percentages inform the minimum number of OGSs required for key servicing over the maximum viewing distance associated with different orbital altitudes if operating in the untrusted node framework. We notice a trade-off between having a large viewable maximum distance and the greater losses incurred when operating from higher altitudes. Smaller altitudes generate higher keys for small OGS separations, but with diminishing returns at higher $d$, since both optical links from a smaller altitude satellite have shallow angles that pass through a greater portion of the atmosphere, compared with higher altitude satellite links, together with a reduced simultaneous contact time. This effect is observed through a steeper drop-off in key rate for lower altitudes. Additionally, if key generation extends to larger baselines, $d$, a crossover in the key rate performance of two different altitudes occurs.

As the OGS separation $d$ increases, higher channel loss generally reduces the number of detected photon pairs, exacerbating statistical fluctuations. The optimisation parameters respond accordingly to address a delicate interplay between estimated error functions and security of finite key generation. Generally, the tolerated error rate $\delta$ increases with distance to accommodate larger QBERs due to reduced signal-to-noise ratio. The sampling ratio $\beta$, governing the raw key fraction used for parameter estimation, increases to counter larger statistical fluctuations in the measured error rate. The excess error rate above threshold, $\nu$ also increases with distance, reflecting a growing gap between observed and true error rates in low-count regimes, with the auxiliary parameter $\xi$ also increasing to preserve a tight finite-sample bound. These variations in the optimisation parameter enables the protocol to dynamically maximise key generation subject to security constraints across different satellite altitudes and losses. An illustration of how the parameters vary with increasing $d$ is included in Supplementary Figures (Fig.~\ref{fig:opt_param}).

As a point of comparison, the Micius satellite performed dual-downlink BBM92 from an altitude of 500~km between the Delingha and Nanshan ground stations, 1120~km apart. With the chosen physical parameters of our model and $h = 500$~km, our model was able to reach similar performance at $d = 1100$~km, the largest value achieved on Fig.~\ref{fig:system_geom}(b). Micius operated with a lower source rate, $f_s=5.9$~M pairs s$^{-1}$, higher $\text{QBER}_\text{I}$, though with significantly larger $T_\mathsf{X}$ and $R_\mathsf{X}$ values of 13~cm and 1.2~m, respectively. With our model adjusted to Micius' source rate and transmitter and receiver dimensions, but maintaining all other parameters as in Table~\ref{tab:system_parameters}, including $\text{QBER}_\text{I}=0.5\%$, our simulation gives an expected key length of $32.7$~kbits for a single overpass (red mark on Fig.~\ref{fig:system_geom}(b)). Despite a much lower source rate, the key rate still performs slightly better with Micius' larger apertures. This highlights the importance of designing missions with reduced system loss, with diffraction losses accounting for the largest contribution. Specifically, a satellite with diffraction parameters matching those of Micius is capable of generating keys at altitudes and OGS separation distances beyond the limits illustrated in Fig.~\ref{fig:system_geom}. Our baseline system parameters are smaller than Micius' to reflect recent missions for global quantum networks that favour constellations of small satellites~\cite{sidhu2021advances,belenchia2021quantum}, owing to improved flexibility of deployment. Other system parameters, such as source rate, must be increased in order to overcome the much higher diffraction losses caused by smaller telescope apertures. 

In reality, the Micius satellite reported lower key lengths, with a final bit string of the order of $10^2$ after multiple overpasses~\cite{Yin2020_N}. There may be varying losses and other parameter assumptions in our model that do not align with Micius which would give this discrepancy in results. We adjusted our model's source rate, diffraction parameters, the satellite elevation and OGS separation, whilst maintaining the rest of the values given in Table \ref{tab:system_parameters}. For example, the geometry of Micius' overpass geometry of the OGSs are significantly different from an ideal zenith-zenith configuration, thus is in a higher loss regime than assumed in the results presented in Fig.~\ref{fig:system_geom}. We have chosen the zenith-zenith overpass, the lowest loss case with a symmetric overpass directly passing over both ground stations, whereas Micius had asymmetric overpasses with minimum link lengths on each channel being greater than the satellite altitude. Supplementary Fig.~\ref{fig:grid} demonstrates how the overpass loss profile is impacted by variations in the orbit.


\subsection{Overpass geometry and annual SKL}
\label{subsec:annual_SKL}

\begin{figure*}[t!]
    \centering
    \includegraphics[width=1\linewidth]{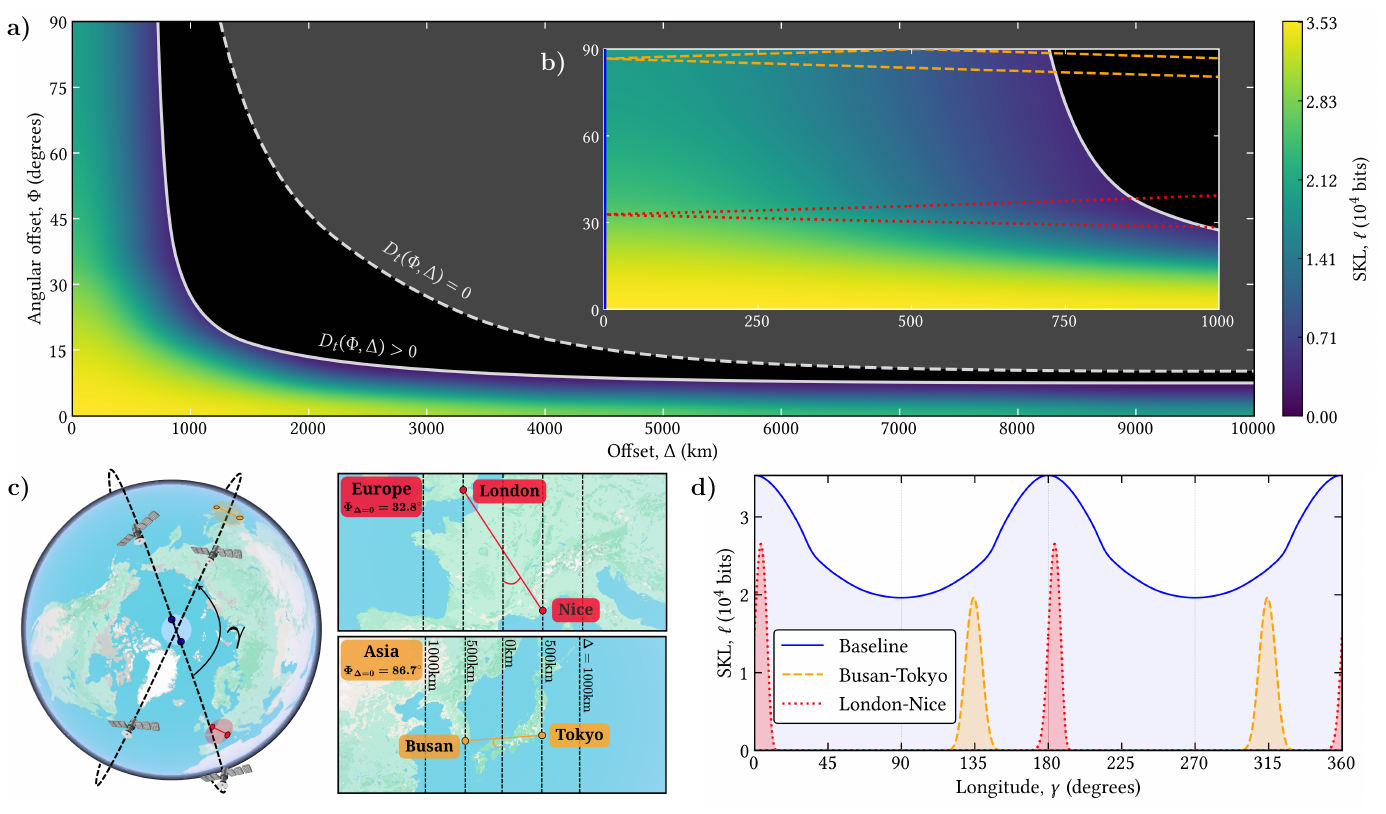}
    \caption{\textbf{Performance under varying overpass geometries}. Parameters: $d=1000$~km, $h=500$~km. \textbf{a)} Optimised SKL across the overpass parameter space defined by the angular offset $\Phi$ and the distance $\Delta$ between the satellite ground-track intersection and the midpoint of the OGS baseline. Only the range $\Delta\in[0,\pi R/2]$ is shown, beyond which the geometry is symmetric. Regions with vanishing optimised SKL due to excessive loss and absence of coincident detections are bounded by the solid and dashed white lines, respectively. The longitudinal coordinate $\gamma$ parameterises a trajectory $(\Delta(\gamma), \Phi(\gamma))$ through this parameter space over the course of a year. Since only a restricted region of $\Delta$ supports secure key generation, inset \textbf{b)} shows a magnified view of the key-generating region, together with the trajectories corresponding to the OGS pairs illustrated in \textbf{c)}, each with $d=1000$~km. The symmetric baseline configuration ($\Delta=0$, $\Phi(\gamma)=\gamma$) is shown in blue, while the red and orange trajectories correspond to more realistic OGS site locations in Europe and Asia, namely London-Nice and Busan-Tokyo pairings, respectively. Map data provided by Google (INEGI, TMap Mobility). \textbf{d)} As the Earth rotates, $\gamma$ is uniformly sampled, tracing points along these trajectories. The resulting per-pass optimised SKL is illustrated as a function of $\gamma$, with shaded areas yielding the annual SKL through Eq.~\eqref{eqn:annual_SKL_long}. System parameters are given in Table~\ref{tab:system_parameters}.}
    \label{fig:annual_SKL}
\end{figure*}

\noindent
We now examine the finite SKL for more general overpass geometries for a given $d$ and $h$. As the Earth rotates, each overpass samples different $(\Phi,\Delta)$ values, each determined by the OGS locations and the orbital inclination of the satellite and yielding different optimised SKL values. However, as illustrated in Fig.~\ref{fig:annual_SKL}(a), the non-zero optimised key generation region is confined to a small region of the $(\Phi,\Delta)$ parameter space (highlighted in inset \textbf{b)}), reflecting the stringent conditions required for successful key generation. This region is bounded by two distinct limits governed by different physical mechanisms. The lower solid white line marks a loss-limited boundary, defined implicitly by $\text{SKL}(\Phi, \Delta)=0$ with $D_t(\Phi, \Delta)>0$. It marks the maximum tolerable loss for a given geometry, above which the coincidence rate is suppressed to a level where the SKL vanishes. The upper dashed white line marks a geometry-limited boundary where no coincident detections are observed, $D_t(\Phi, \Delta)=0$, due to the absence of simultaneous visibility of both OGSs. Together, these boundaries delineate the narrow region in which key generation is possible.

Within the key-generating region, the SKL is maximised when both $\Phi$ and $\Delta$ are small, corresponding to near-overhead passes with minimal geometric and atmospheric losses. It decreases monotonically as either parameter increases due to longer link distances and increased atmospheric path length, which reduce the detected coincidence rate. The optimised SKL exhibits greater sensitivity to $\Phi$ than to $\Delta$ reflecting the compounded impact of both enhanced channel losses and the rapid reduction in overpass duration during which both OGSs are simultaneously visible to the satellite. In contrast, moderate increases in $\Delta$ at fixed $\Phi$ primarily extend the link distance while maintaining relatively favourable elevation angles, resulting in a more gradual degradation of the SKL. This asymmetry highlights how even modest deviations from zenith trajectories can substantially reduce the secret key yield, indicating optimal performance when the OGS baseline is aligned with the orbital inclination. Beyond the key-generating region, the overpass geometry exhibits a symmetry about $\Delta' = \pi R/2 \approx 10,000$~km, beyond which the roles of the two OGSs are exchanged as the satellite ground track moves away from one OGS and towards the other. Consequently, both the boundaries and the key-generating region are mirrored about $\Delta'$.

To account for the full distribution of satellite overpasses, the expected annual SKL provides a more meaningful performance metric for SatQKD systems than the SKL obtained from any single overpass. The concept of an annually averaged SKL as a performance metric was introduced in Ref.~\cite{Sidhu2022_npjQI}. Here, we extend that methodology to our configuration involving two OGSs, which can more readily be generalised to additional satellite-OGS network geometries. Precisely computing the annual SKL would require a full orbit-by-orbit simulation over a year, but is sensitive to initial conditions and orbital perturbations. In order to easily derive trends and qualitative understanding we choose an alternative sampling that permits an analytic distribution to efficiently compute the annual SKL.

As in the single satellite and single OGS case, we can estimate the annual SKL by taking the uniform random sampling of the overpass geometry for a given OGS pair location and satellite orbit. As an example, we assume a near-polar \emph{sun-synchronous orbit} (SSO) so that throughout a year it evenly crosses the entire range of longitudes, denoted $\gamma$. Each value of $\gamma$ uniquely maps to a point $(\Delta(\gamma), \Phi(\gamma))$ in the full overpass geometry parameter space. As the Earth rotates, $\gamma$ is uniformly sampled over the year, corresponding to a locus of points through the $\{\Delta,\Phi\}$ parameter space. The annual SKL can be estimated by averaging the per-pass SKL along the line in the $\{\Delta,\Phi\}$ plane, taking care to scale the line element according to the variation of $\Delta$ and $\Phi$ with respect to $\gamma$. Hence the average SKL can be \je{approximated as}
\begin{align}
\text{SKL}_\text{year} = \frac{N^\text{orbit}_\text{year}}{2\pi} \int_0^{2\pi} d\gamma \; \text{SKL}\left[ \Delta(\gamma), \Phi(\gamma)\right]\, ,
\label{eqn:annual_SKL_long}
\end{align}
where $\smash{N^\text{orbit}_\text{year}}$ is the number of satellite orbits per year. The normalisation factor $2\pi$ emerges from the uniform distribution of $\gamma$. In the general case, the line integral should be over the entire range of $\Phi\in(0,\pi]$ and $\Delta$ taking on both positive and negative values (either OGS is closer to the intersection point). If daylight operation is not assumed, the range of SKL generating overpasses can be restricted to those overpasses occurring for only specific ranges of $\gamma$. To discard daylight operation, we include an additonal factor of $1/2$ to the $\text{SKL}_\text{year}$. Note, while we found daylight operation to be feasible in section~\ref{subsec:block_size}, the reduced performance means we did not include it in our evaluation. Therefore the values for $\text{SKL}_\text{year}$ reported here represent a lower bound on the performance.

We first consider the symmetric case where the OGSs are arranged about the North pole and the satellite follows a polar orbit, for which the overpasses are characterised by $\Delta=0$~km and $\gamma=\Phi$. This corresponds to the vertical red line in inset Fig.~\ref{fig:annual_SKL}(b), defining the trajectory sampled by the baseline system through the overpass parameter space over the course of a year. While this symmetric configuration provides a useful upper bound, realistic OGS deployments will deviate from this idealised geometry. To capture such scenarios, we consider two additional representative OGS pairings separated by approximately 1000~km, each sited at existing or planned metropolitan-scale quantum networks. Specifically, we examine the \emph{London-Nice} (LN) and \emph{Busan-Tokyo} (BT) links, illustrated in Fig.~\ref{fig:annual_SKL}(c), which exhibit markedly different overpass geometries. At $\Delta=0$~km, these links correspond to angular offsets of $\Phi^\text{LN}_{\Delta=0}=32.8^\circ$ and $\Phi^\text{BT}_{\Delta=0}=86.7^\circ$, respectively. These examples explore instances of relatively low and high angular deviation, tracing distinct trajectories through the $(\Phi,\Delta)$ parameter space as illustrated in inset Fig.~\ref{fig:annual_SKL}(b).

Qualitatively, since practical baselines will not pass through the North Pole, both $\Delta(\gamma)$ and $\Phi(\gamma)$ trace more general trajectories through the entire parameter space (Supplementary Fig.~\ref{fig:metropolitan_trajectories}), leading to reduced annual SKLs. To rigorously quantify the impact of site location on the long-term key generation performance, Fig.~\ref{fig:annual_SKL}(d) illustrates the optimised SKL evaluated along all three trajectories as a function of longitude $\gamma$. For the baseline configuration, the SKL decreases monotonically with increasing overpass angles $\Phi \in [0, 90^\circ]$ due to increasing channel loss and reduced overpass duration. When $\Phi=0^\circ$, the satellite ground track aligns with the OGS baseline, yielding minimal geometric loss and maximal key generation. As $\gamma$ (and hence $\Phi$) increases, the overlap in visibility between the two OGSs diminishes and the effective QBER rises, leading to a reduction in the per-pass finite key yield. Owing to the baseline geometry symmetry, this behaviour repeats for $\Phi>90^\circ$.

In contrast, the LN and BT configurations exhibit more restricted regions of non-zero SKL, as a consequence of their off-axis geometries. Both support key generation only over a narrow range of $\gamma$. The two peaks observed for each metropolitan configuration arise from the symmetry of the parameter space about $\Delta'$. Longitudes separated by $180^\circ$ correspond to equivalent overpass geometries with the roles of the two OGSs exchanged, causing each viable key-generating configuration to appear twice in Fig.~\ref{fig:annual_SKL}(d). The shaded areas under each curve quantify the annual SKL via Eq.~\eqref{eqn:annual_SKL_long}. Accounting for the half factor to accumulate key only during night time operation, we obtain $\text{SKL}_\text{year}=70.10$~Mb for the baseline scenario, compared with $\text{SKL}^\text{LN}_\text{year}=3.55$~Mb and $\text{SKL}^\text{BT}_\text{year}=3.65$~Mb for the metropolitan combinations. The relative reduction in annual key accumulation of the metropolitan pairs reflects their less favourable overpass geometries and highlights the strong dependence of long-term performance on geographic configuration. Further, despite the requirement for coincident detections events in entanglement-based QKD, the metropolitan link performances remain competitive with those reported for the DS-WCP BB84 protocol operating with a total system loss of 40~dB~\cite{Sidhu2022_npjQI}. 

Before concluding, we note that the framework developed here can be readily applied to evaluate the average annual key for arbitrary OGS placements, enabling further quantitative assessment and optimisation of satellite QKD service provision between general metropolitan locations.


\subsection{Enabling technology platforms}
\label{subsec:space_comp_ent_sources}

\noindent
This section will outline practical developments in the necessary components that are required to enable secret key generation at the rates presented in the previous sections. Developing \emph{entangled photon sources} (EPS) which are compatible with deployment on satellites remains an active challenge in the realisation of entanglement-based SatQKD. Satellite deployment requires engineering of these sensitive quantum payloads to maintain stable performance within a challenging environment. This is exacerbated when considering small satellites, like CubeSats, which have very restrictive SWaP constraints. High pair generation rates are essential for this application, informing our chosen source rate of 100~Mpairs~s$^{-1}$, a rate that reasonably reflects the progression in this area~\cite{Anwar2021}. The demonstrations of entanglement on board satellites have both generated the photon pairs using SPDC within interferometers that give polarisation Bell states~\cite{Yin2020_N, Villar2020}. The Sagnac loop is a widely implemented source architecture for in-lab applications due to it's ability to achieve high fidelities~\cite{Chapman:20}. It requires a large footprint relative to other sources, however this was feasible for use on the Micius satellite due to its large size. For the demonstration of entanglement on a nanosatellite by the SpooQy-1 mission, a parallel crystal design was implemented, which using two non-linear crystals to generate the components of the Bell state~\cite{Villar_2018}. This is a far more compact approach, however may have issues with stability and requires uniformity between the two non-linear crystals.

Linear displacement interferometers have been proposed as an alternative that allows for the benefits of a sagnac within a much smaller footprint~\cite{Lohrmann2020, Anwar2022}. This source was chosen for deployment on the recently launched SpeQtre satellite~\cite{sivasankaran2022}, a 12U CubeSat and follow up mission to SpooQy-1, aiming to also demonstrate satellite-to-ground communication. Variations on this source giving benefits in compactness, stability and source brightness have also been proposed for their suitability for use on compact platforms~\cite{Fazili2024, McCarthy:25}. These are all CW-pumped sources of entangled photons. Higher rates of photons can be generated more efficiently with a pulsed pump due to the high pulse peak powers~\cite{Li_2015}, however this does increase the likelihood of emitted pairs with photon numbers $>$1. A CW-pumped source was modelled in this work to remove the security implications of multipair emission, as is discussed in Methods~\ref{subsubsec:source}. Wavelength multiplexing is another technique we rely on in our model. It can be used to maximise secret key lengths by allowing for multiple detection channels~\cite{Pseiner_2021} or networks with multiple users~\cite{Fan2025}.

In order to avoid the challenges of deploying the entanglement source on the satellite, the Canadian QEYSSAT mission is opting for a ground based EPS, with a single uplink and fibre channel~\cite{Jennewein2014_SPIE}, utilising non-degenerate photon pair wavelengths to limit losses in the fibre and free space channels~\cite{Jaeken_2025}. However, uplinks tend to be higher loss and there remains the distance limitation imposed by the fibre link on one arm. 

Integrated sources of entangled photons are another solution providing reliable and compact sources for satellite deployment~\cite{Wang_2019}. There is greatest maturity in silicon photonic integrated circuits, however these only have $\chi-3$ non-linearity so rely on the less efficient process of \emph{spontaneous four-wave mixing} (SFWM). Lithium niobate is an alternative material often used for integrated optics with $\chi-2$ interaction, allowing for SPDC. Though on-chip non-linear processes can be more efficient than using bulk optics, interfacing the light on-chip with other optical components can be lossy on either platform. The development of integrated sources in the NIR is far less mature than it is for 1550~nm. Although 1550~nm sources have clear applications when discussing fibre networks, the greater diffraction effects in free-space links may make useful key lengths more challenging to achieve.

Enhanced single photon detection will also improve the achievable key generation and potentially allow for longer OGS separations to become feasible. Many Si-SPADs have detection efficiencies up to 70\% for the wavelength considered~\cite{hoque2025}. Though typically designed for the C-band, \emph{superconducting nanowire single photon detectors} (SNSPDs) also operate for shorter wavelengths and report detection efficiencies of $>$90\%~\cite{You17}. Alongside enhanced detection efficiency, SNSPDs also offer very low timing jitter which will enable higher detection rates by allowing faster processing of incident photons. \emph{Photon number resolving} (PNR) technology has the ability to identify multiphoton detection events~\cite{Los2024}, and so can be removed from the block. Though this is unlikely with cw-pumped sources, there is a higher probability of occurrence with pulsed sources, which have the ability to generate higher pair production rates. Therefore PNR detection would allow for use of pulsed sources without the security vulnerability of multiphoton events.


\section{Discussion}
\label{sec:conc}

\noindent
Satellite-based entanglement distribution provides a cornerstone capability for global-scale quantum networking, enabling secure communications and coordinated quantum operations beyond the limits of terrestrial fibre networks. SatQKD is a precursor application, using key rates as a benchmark to develop the underlying architecture that will power more general applications, such as distributed quantum computing, sensing, and communications applications. Entanglement-based SatQKD offering fundamental security and the untrusted-node communication networks. However, despite the architectural importance, a quantitative understanding of untrusted-node SatQKD under realistic operational constraints has remained limited. This work addresses this gap by developing a comprehensive end-to-end model of a dual-downlink BBM92 satellite QKD system and applying a rigorous finite-key security analysis to assess its achievable performance across realistic network geometries and operating conditions.

Using this framework, we examine the finite-key performance of the protocol. With two spatially separated OGSs, the overpass loss profile is generally asymmetric, rendering conventional time-window-based block selection suboptimal. We introduce an alternative threshold-based block construction scheme that selects data below a QBER threshold, making our approach (1) independent of any symmetry in the overpass loss profile, (2) readily applicable to general network topologies, and (3) adaptable to varying channel conditions to enable improved performance. Applying this optimisation, we first investigate how increasing background illumination affects the viability of untrusted-node SatQKD. During night-time operation, the optimal strategy is to use the entire overpass data, yielding a secure key length that increases linearly with block size. As the background level increases, we identify a broad regime spanning astronomical, nautical, and civil twilight, extending into early clear daytime conditions, in which secure key generation remains possible without fundamental protocol changes using baseline system parameters representative of current technologies. In this regime, admitting all available data becomes detrimental, as high-QBER events increasingly dominate the finite-key statistics. Instead, joint optimisation over both the block size and protocol parameters is required, revealing a fundamental trade-off between data volume and data quality in the finite-resource regime. The persistence of non-zero finite keys into early daytime conditions demonstrates a substantial operational window beyond conventional night-time regimes typically considered in the literature. Further tolerance to elevated background counts could be achieved through modest improvements to experimentally accessible system parameters and background mitigation strategies, providing a practical pathway towards more robust daytime operation. A comparison with P\&M protocols reveals that although BBM92 generally yields lower SKL than DS-BB84 under comparable conditions, it exhibits greater resilience to channel loss.

Second, we use our framework to address a pressing architectural question for entanglement-based satellite QKD: \emph{how far can two OGSs be separated while still enabling key generation via BBM92?} To answer this, we quantify how the finite SKL depends on the geometric configuration of the satellite-OGS system. Increasing the satellite altitude increases diffraction and propagation losses, while increasing OGS separation introduces additional atmospheric loss through larger slant angles. Together, these effects compound to limit key generation and, more broadly, quantum networking applications, highlighting a clear trade-off between coverage area and performance. We rigorously quantify this trade-off across a range of satellite altitudes representative of near-term mission proposals. For example, a typical LEO orbit at 500~km can only generate keys when the two OGSs span approximately 23\% of the satellite's maximum viewable distance. Despite their reduced viewing distances, VLEOs generally provide the highest key lengths for OGS separations up to 1,300~km. The development and adoption of air-breathing propulsion systems to enable long duration missions in VLEO will therefore be essential to realise these advantages. More generally, these results emphasise the need to co-design network constellations and OGS placement with protocol-level performance. Our framework provides a quantitative toolkit to support this effort. Though this analysis was limited to considering zenith-zenith overpasses, as they present the lowest loss case, we can assume that our results represent an upper bound on the possible range.

Finally, we generalise the definition of the annual SKL to assess system-level performance across general overpass geometries relevant to realistic deployments. We apply this framework to representative OGS pairings aligned with existing and planned metropolitan quantum networks, satisfying $d=1000$~km. We qualitatively and quantitatively characterise the key-generation regime, which occupies a small region of the $(\Delta, \Phi)$ parameter space bounded by both loss- and geometry-imposed constraints. For a polar orbit, the uniformly sampled longitude $\gamma \mapsto (\Delta(\gamma), \Phi(\gamma))$, tracing a trajectory in this space for a given OGS pair. Under optimal alignment, our baseline scenario yields $\text{SKL}_\text{year}=70.10$~Mb, whereas realistic site configurations exhibit a substantial reduction in annual key yield due to the restricted key-generation regime, with $\text{SKL}^\text{LN}_\text{year}=3.55$~Mb and $\text{SKL}^\text{BT}_\text{year}=3.65$~Mb for the metropolitan links. These results highlight the strong dependence of long-term performance on the system geometry, and the importance of co-designing ground station placement with satellite orbits. The annual SKL metric therefore offers a practical tool for network-level design and optimisation, and can be readily applied to any satellite-OGS geometry to determine the sustained communication performance of a quantum network. 

Revisiting the open question of how entanglement-based SatQKD performs under realistic operating conditions, we develop a rigorous simulation framework that quantifies optimised SKL yields across realistic orbital dynamics and network geometries under representative environmental regimes. By connecting our results to earlier finite-key SatQKD studies, we clarify the distinct advantages and limitations of the BBM92 protocol in the finite-resource regime. Our work establishes a broad foundation for further investigations into entanglement-based satellite communications architectures. First, systematic key generation analyses across broader system configurations can directly inform the design and development of emerging OGS research facilities and international missions, such as the HyperSpace initiative aiming to connect Canada and the EU via high-dimensional entanglement~\cite{Kanitschar2024_PRApp}. While our present study focuses on LEO architectures, the framework readily extends to higher-altitude platforms and longer intercontinental baselines that such missions must contend with. Second, our work naturally extends beyond two-node scenarios to multi-OGS, multi-satellite constellations for continuous key delivery. Third, the underlying loss and geometric modelling can be directly integrated with satellite-based quantum repeater architectures, enabling dynamic and accurate estimates of entanglement distribution rates for a range of applications beyond QKD. Collectively, this line of research will help determine the viability and requirements of long-range quantum networking. 

During submission, we noted Ref.~\cite{Wang2026}, which studies dual-downlink entanglement-based satellite QKD between specific ground stations with varying apertures at each site used by the Micius satellite. The work consider factors such as overpass asymmetry and cloud thickness. Together with our generalised approach to model the performance with orbits and ground station locations, providing a complimentary analysis on range limits and effects of different background light conditions.

\appendix

\section*{Methods}
\label{sec:methods}

\noindent
We start by first outlining our key rate model for the dual downlink, finite-key BBM92 protocol. For a system of one satellite and two OGSs, we outline our model in four steps: First, section~\ref{subsec:link_geom} defines the link length and elevation dynamics of a general orbital overpass. Second, given the length, section~\ref{subsec:loss_modelling} sets the baseline transmission probability for signal photons reaching the OGSs, accounting for diffraction, atmosphere, intrinsic system loss. This probability assigns a link efficiency that quantifies what gets through the channel. Third, section~\ref{subsec:ext_counts} collects all noise sources into one framework by accounting for multi-photon pair production, imperfect entanglement fidelity, background counts, dark counts, and afterpulsing. This section quantifies the count contribution at each OGS that arrives but should not, which we coin extraneous counts. Finally, in section~\ref{subsec:block}, we describe how loss and noise combine into secure key metrics. While we describe a one satellite, two OGSs system, these methods readily extend to a more complex system of pair-wise links.


\color{black}
\subsection{Link geometry}
\label{subsec:link_geom}

\noindent
Here, we outline our model of different loss mechanisms to determine the total dual-downlink efficiency.


\subsubsection{Orbit parametrisation}
\label{subsubsec:orbit_parameterisation}

\noindent
We consider a satellite in a circular polar orbit. This trajectory can be modified to model different geometries. In the reference frame of the two OGSs, a general overpass can be parameterised by two angles. The first, $\Phi$, represents a rotation around the Earth's polar axis. Intuitively, this sets the angular offset between the great-circle arc connecting $\text{OGS}_A$ and $\text{OGS}_B$ and the satellite's ground track. When $\Phi=0^\circ$, the satellite ground track runs parallel to the OGS baseline, and orthogonal when $\Phi=90^\circ$. The second, $\Xi$, tilts the orbital plane by a rotation about the intersection of the equator and the prime meridian. This tilt shifts the satellite's ground track by a polar distance
\begin{align}
    d_\Xi = R\left(\frac{\pi\Xi}{180^\circ}\right),
\end{align}
where $R$ is the Earth's radius. For $\Phi=\Xi=0^\circ$, the ground track follows the prime meridian and recovers a standard polar orbit. More generally, $\Xi$ controls the north-south displacement of the orbit, so that, for example, when $\Phi=90^\circ$ and $d_\Xi$ is half the separation between the OGSs, the satellite passes directly over one station while reaching a maximum elevation of $45^\circ$ at the other. Further, we define the orbit offset
\begin{align}
    \Delta = \Abs{\frac{d_\Xi}{\sin\Phi}},
\end{align} 
\je{as a measure of how the geometry favours one OGS over the other}. Figure~\ref{fig:geom} illustrates this offset parameterisation. For large tilts, typically $\Xi \gtrsim 15^\circ$, the overlap of the satellite visibility windows can vanish, eliminating any region of joint visibility for feasible communication.

Taking both rotations collectively, we write the time-dependent coordinates of the satellite
\begin{align}
\begin{split}
r_\text{sat} (t) & = \texttt{rot}(\Phi, \Xi) [0, (R+h)\sin\omega t,  (R+h)\cos\omega t ] \\
    & = 
\begin{pmatrix}
	(R+h)(\cos{\omega t}\cos{\Phi}\sin{\Xi} - \sin{\Phi}\sin{\omega t}) \\
	(R+h)(\cos{\omega t}\sin{\Phi}\cos{\Xi} - \cos{\Phi}\sin{\omega t}) \\
	(R+h)\cos{\omega t} \cos{\Xi}
\end{pmatrix}\, ,
\end{split}
\end{align}
where $h$ the orbital altitude and $\omega$ the orbital angular velocity. While our model permits any arbitrary coordinates for the OGS locations, we assume the OGS coordinates are symmetrically displaced from the North Pole by an angle $\alpha$, corresponding to an OGS separation distance, $d$. 


\subsubsection{Link length and elevation angle}
\label{subsec:link_length}

\noindent
From the satellite's dynamic coordinate, the link range and elevation between each OGS is $l^i_t = \abs{r_\text{sat}(t) - r^i}$ and
\begin{align}
    \sin(\theta^ i_t) = \frac{z_\text{sat}(t) - z^i}{h},
\end{align} 
respectively for $i = \{A, B\}$. We compute both quantities for the overpass time that falls within the visible region, defined as the overpass time duration where the satellite's elevation satisfies $\theta^{A/B} \geq \theta_\text{min} = 10^\circ$ for both OGSs. This constraint ensures simultaneous communication with both ground stations, which will later help to determine the coincident clicks. Each quantity is evaluated at one-second intervals during the entire visible region. For a representative altitude of 500~km, the duration of joint visibility between the two OGSs ranges between 250 to 500~s, depending heavily on the overpass geometry.


\subsection{Loss modelling}
\label{subsec:loss_modelling}

\noindent
From the satellites instantaneous coordinate, we determine the loss associated with the quantum channel between each OGS. Denoting the channel loss for the satellite-$\text{OGS}_A$ link as $\eta_t^A$ and the satellite-$\text{OGS}_B$ link as $\eta_t^B$, then the total system loss is $\eta_t = \eta_t^A\eta_t^B$. The losses associated with each link are separated into three distinct linear loss contributions,
\begin{align}
\eta_t^i = \eta_\text{diff}(\lambda, \theta^i_t) \, \eta_\text{atm}(\lambda, \theta^i_t) \, \eta_\text{int}\, ,
\label{eqn:loss_contr}
\end{align}
for $i \in \{A, B\}$ and where $\eta_\text{diff}$ defines losses from diffraction effects, $\eta_\text{atm}$ from atmosphere effects that include scattering and absorption, and $\eta_\text{int}$ defines a fixed elevation-independent intrinsic system efficiency corresponding to internal losses, and beam misalignment. Eq.~\ref{eqn:loss_contr} provides a general approach to modelling losses for any SatQKD system, with time (or equivalently elevation) dependent losses and losses intrinsic to the system. Details for each loss contribution is given in the following sections to determine the combined system loss.


\subsubsection{Diffraction losses}
\label{subsubsec:diff_loss}

\noindent
A dominant contribution to loss for each channel is diffraction, which broadens the beam during signal propagation from the satellite to the OGS. The extent of beam broadening depends on a number of factors, including the satellite transmitter aperture $T_\mathsf{X}$, source wavelength $\lambda$, and the channel range $R(t)$. We estimate diffraction losses by calculating the far-field Fraunhofer diffraction of an initially truncated Gaussian field distribution with a beam waist of $w_0$ at the transmission aperture. To then compute the probability that a single photon exiting the transmit aperture is collected at receiver aperture, we take the ratio of the integrated power density across the transmitter aperture, $P_T$, and the receiver aperture, $P_R$ such that, in dB,
\begin{align}
  \eta_\text{diff}^\text{dB}\left(\lambda,\theta^i_t\right) = -10 \log_{10}\left(\frac{P_R}{P_T}\right)\, .
\end{align}
This approach determines the efficiency of each channel due to beam broadening. Due to the single photon nature of the output of SPDC sources, we optimise the beam waist to be in the order of the transmitter aperture diameter, $w_0 = T_\mathsf{X}/2$. The impact of a central beam obscuration due to secondary mirrors typical of Cassegrain-type reflecting telescopes could be considered~\cite{Bourgoin:2013fk} but has no significant impact on the analysis.


\subsubsection{Atmospheric attenuation}
\label{subsubsec:atm_loss}

\noindent
An important contributor to the losses in a satellite-to-ground system is the attenuation due to interaction with the atmosphere, causing scattering and absorption. The data for the atmospheric attenuation is extracted from MODTRAN, which provides an elevation-dependent efficiency of the link. This is taken for a range of wavelength within the NIR region, allowing for flexibility in the degeneracy of the photon pairs. The linear transmissivity, $T^i_{\lambda}(t)$, for link $i$ is taken for $0^\circ \leq \theta \leq 90^\circ$, at 3$^\circ$ increments. This data is interpolated using \texttt{scipy.interp1d}. This allows us a full data set from which to access the transmissivity for all elevation angles calculated by the orbit geometry. The efficiency of the link is converted into the loss contribution, in dB,
\begin{align}
  \eta_\text{atm}^{\text{dB}}\left(\lambda,\theta^i_t\right) = -10 \log_{10}\left(T^i_{\lambda} (t)\right),
\end{align}
for signal wavelength $\lambda$ and the time $t$ during the satellite's overpass.

\begin{figure}
    \centering
    \includegraphics[width=\columnwidth]{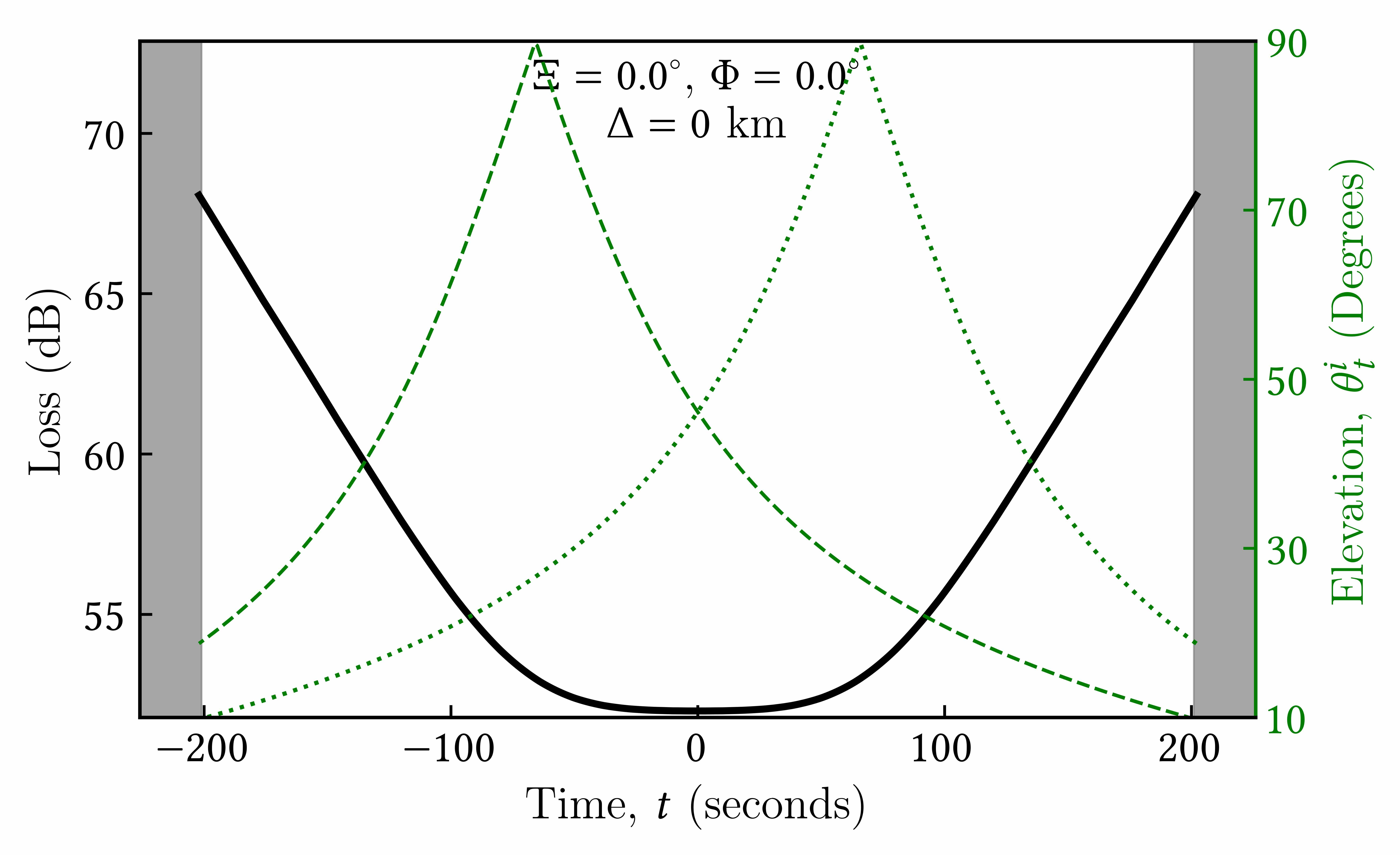}
    \caption{\textbf{Link model for downlink channel}. Joint loss and elevation as a function of time for a zenith overpass ($\Xi=\Phi=0^\circ$). The black line illustrates the total system loss as a function of time, with the green dashed and dotted lines indicated the elevation for link A and B respectively. The grey region marks time windows where the satellite’s elevation for either link drops below $10^\circ$. System parameters as in Table~\ref{tab:system_parameters}. Additional loss profiles are provided in Fig. \ref{fig:grid} in Supplementary Figures.}
    \label{fig:single_geom_loss}
\end{figure}%

\color{black}


\subsubsection{`Intrinsic' system loss}
\label{subsubsec:intrinsic_loss}

\noindent
Intrinsic losses are loss contributions independent of the satellite and OGS geometries, and are therefore taken as constant throughout the visible time window of both OGSs. We consider a number of individual contributors to the intrinsic system loss and assign each contribution with 50\% per downlink channel, or 3~dB. First, detection losses accounts for the efficiency of Si-SPADs. Second and third are losses associated with the receiver and collection optics respectively. Finally, the pointing and tracking losses are combined to account for the inefficiency of the adaptive beam steering optics. To benchmark our results with empirically available results from Micius, we add an additional 1.5~dB loss per link. Taken collectively, we therefore take an intrinsic loss of $\eta_\text{int}^{\text{dB}} = 13.5~\mathrm{dB}$ for each link. Therefore, the total intrinsic loss for the dual down link architecture is 27~dB.

Taken collectively, Fig.~\ref{fig:single_geom_loss} illustrates the time-dependent elevation of each link, $\theta_t^i$, and the total system loss, $\eta_t$ for the nominal set of system parameters in Table~\ref{tab:system_parameters}. An extended data set for a range of different satellite trajectories is shown in Extended data 1.


\subsection{Extraneous counts}
\label{subsec:ext_counts}

\noindent
Errors in key generation arise from counts that do not originate from a single source pair of entangled photons. These errors can be related to the source (multi-photon pair production, imperfect entanglement fidelity), detector imperfections (dark counts and afterpulsing), or 
background light collected by the ground station telescopes.


\subsubsection{Source model}
\label{subsubsec:source}

\noindent
We consider a \emph{continuous wave} (CW)-pumped spontaneous parametric down-conversion source of polarisation entangled photons, which have been demonstrated for satellite QKD missions~\cite{Yin2017_S, Villar2020}. In a \emph{spontaneous parametric down-conversion} (SPDC) source of entangled photons, a pump photon enters a second-order nonlinear medium and converts into two daughter photons, called signal and idler, such that the sum of the signal and idler frequencies is equal to the pump frequency. This process can be degenerate, where the signal and idler frequencies are equal with $\omega_\text{signal} = \omega_\text{idler} = \frac{1}{2}\omega_\text{pump}$, or non-degenerate with $\omega_\text{signal} \neq \omega_\text{idler}$. The model is able to consider either case as the wavelength-dependent considerations are taken separately for each OGS link. 

An important aspect of accurately modelling an SPDC source is the treatment of multiple photon pair emissions, which can introduce uncorrelated detection events and additional errors. Multiple entangled pairs emitted within the same coincidence window may still be detected. With no way to distinguish which photons belong to which pair, this scenario would increase the polarisation error in the raw key. While \emph{prepare-and-measure} (P\&M) protocols based on \emph{weak coherent pulse} (WCP) sources are vulnerable to 
\je{instances of} \emph{photon-number-splitting} (PNS) attacks, entanglement-based (EB) protocols (with the entangled source regarded as part of the untrusted channel) make no assumptions about photon number distribution of the emitted states. This distinction gives EB-protocols an inherent robustness against PNS attacks. However, the security proof does depend upon only a single photon coincident detection events at each OGS, otherwise multi-photon eavesdropping attacks become possible \cite{Lutkenhaus:1999,gottesman2004security}. With no decoy-protocol available this condition can be empirically monitored by implementing a \emph{squashing model} \cite{Beaudry:2008,Gittsovich:2014}. In practice this requires that whenever Alice or Bob observe a double-click event, they record a random bit value for that round. In this way, any multi-photon attacks are accounted for in the QBER estimation. Here, these contributions are included in the intrinsic QBER coefficient.

Formally, the probability of generating $n$ photon pairs in a single emission is given by
\begin{align}
    P(n,r) = \frac{(\tanh r)^{2n}}{(\cosh r)^2},
\end{align}
where $r$ is the squeezing parameter for a two-mode squeezed vacuum state generated by the SPDC source. The cumulative probability of emitting more than one pair is $P(n>1, r) = 1 -P(0,r)-P(1,r)$~\cite{Schneeloch_2019}. In practice, $r$ is kept sufficiently small such that $P(n>1,r)$ is negligible compared with other sources of error, including detector dark counts, background light, and optical misalignment. For instance, typical operating values of $r \lesssim 10^{-2}$ correspond to multi-pair probabilities well below $10^{-3}$, which is orders of magnitude smaller than 
other error contributions. CW pumping further suppresses multi-pair generation compared with pulsed pumping, with very high peak powers~\cite{Li_2015}, while maintaining high single-pair rates, making it particularly advantageous for satellite applications where link losses are high. With typical pump powers of a few milliWatts, the ratio of multipair to single 
pair events remain of the order of magnitude of $10^{-10}$~\cite{Schneeloch_2019}. 

Given that the diminishing contribution of multi-pair events, we do not explicitly include them in our modelling. For sources where this contribution is non-negligible, the effect of multi-photon pairs can be readily captured by weighting the coincidence statistics with the probabilities $P(n, r)$. In this case, multi-pair emissions would increase the polarisation error by approximately $P(n>1,r)/2$, reflecting the fact that in half of the coincidence detections the photons would be correctly paired, while the rest would be mismatched, with half the mismatched cases leading to a incorrect result. In the security analysis, this additional polarisation error would be absorbed into the intrinsic QBER term, effectively raising the baseline error budget. Even for conservative estimates of $r$, however, $P(n>1,r)$ remains well below $10^{-3}$, so the resulting correction is negligible compared with other dominant noise sources in satellite-based scenarios.

Although true multi-pair production events are expected to be rare, at sufficiently high pair-production rates it is not possible to distinguish independent pair-generation events within a single coincidence window. In this regime, if the pair-generation rate exceeds the rate of detection windows, multiple independent pairs can arrive within the same detection window, which induces the same effect as a multi-pair generation event. The associated error contribution is given by $f_s\Delta\tau/4$, where $f_s$ is the pair-generation rate and $\Delta\tau$ the coincidence-window duration. Using the parameters in Table~\ref{tab:system_parameters}, this corresponds to an error of approximately $1.25\%$. However, assuming a spectral filtering bandwidth of 10~nm, it is reasonable to consider frequency multiplexing across 10 channels, reducing the effective contribution to $0.125\%$. This residual error can therefore be absorbed into the intrinsic error value $\text{QBER}_\text{I}$.

Finally, we account for imperfect Bell-state generation in the SPDC source itself, which contributes an intrinsic polarisation error. We model this by assuming a Bell-state fidelity below unity and assign a QBER contribution of 0.5\%. Together with the extraneous count mechanisms described in subsequent subsections, this defines the baseline error floor used in our finite-key security analysis.


\subsubsection{Background light}
\label{subsubsec:background}

\noindent
\je{In such approaches, one cannot neglect the influence of background light.}
The contribution of background light to the error budget is estimated using radiance data obtained from MODTRAN together with the parameters of the receiver system. We evaluate $p_\text{bg}$, and hence its contribution to the QBER, independently for $\text{OGS}_A$ and $\text{OGS}_B$ to capture site-specific variations in background radiance, wavelength dependence, and receiver characteristics. Specifically, for a given source wavelength, date and time, we extract the total radiance $H_\text{tot}$ at each ground station location. The radiance, expressed in units of Wcm$^{-2}$~sr$^{-1}$~nm$^{-1}$, represents the power per nanometre wavelength band incident on the receiver area per steradian, and includes both natural (e.g., moonlight, starlight, airglow) and artificial light sources. We then determine the corresponding probability of detecting a background photon at the OGS $i \in \{A,B\}$~\cite{Bourgoin:2013fk}
\begin{align}
p^i_{\text{bg}, t} = \frac{\Delta\tau}{E_\nu}\left\{H^i_\text{tot}(t) \, \pi \, \left( \frac{R_\mathsf{X}^i}{2} \right)^2 \, \pi (\text{FOV}^i)^2 \Delta\lambda^i\right\}\, ,
\label{eqn:bg_prob}
\end{align}
where $\Delta\tau$ is the coincidence window, $E_\nu$ the photon energy, $R_\mathsf{X}$ the receiver diameter, FOV the angular field of view, and $\Delta\lambda$ the spectral filter bandwidth. Note that we have highlighted the time dependence owing to the satellite's dynamic position. Eq.~\eqref{eqn:bg_prob} accounts for the spectral, temporal, and spatial filtering properties of the receiver system. Unless otherwise stated, we assume night-time operation where background contributions are minimised, though our model can readily incorporate daytime or twilight conditions for performance comparison.


\subsubsection{Extraneous detector counts}
\label{subsubsec:detect}

\noindent
Single-photon detectors contribute spurious counts via two dominant mechanisms, dark counts, $p_\text{dc}$, and afterpulsing, $p_\text{ap}$. Dark counts arise from intrinsic noise processes within the detector and are independent of the incident photon flux. By contrast, afterpulsing results from residual charge carriers generated during a detection event, subsequently triggering spurious detections events. The afterpulse probability is therefore proportional to the rate of true detection events and is indirectly influenced by both the source brightness and the background light level. For each channel, we take representative values of $p_\text{dc}=5\times10^{-7}$ and $p_\text{ap}=1\times10^{-3}$.

Cumulatively, the probability that the detector registers an extraneous click from either background light (time-dependent) or dark counts (time-independent) in a given time window is therefore
\begin{align}
p_{\text{ec}, t}^i = p_\text{dc} + p_{\text{bg},t}^i - p_\text{dc} p_{\text{bg},t}^i\, , 
\label{eqn:ec_prob2}
\end{align}
for link $i \in \{A,B\}$ and with $p_{\text{bg},t}^i$ defined in Eq.~\eqref{eqn:bg_prob}. Note this probability accounts for at least one of these two spurious counts mechanisms, with afterpulse effects considered later.


\subsection{Block size and QBER}
\label{subsec:block}

\noindent
The block size is the accumulation of coincident clicks at $\text{OGS}_A$ and $\text{OGS}_B$ within the visible region ($\{\theta_A,\theta_B\}\geq 10^\degree$). From the link efficiencies $\eta_t^A$ and $\eta_t^B$, we compute first the click rate $D_t$, which quantifies the average number of simultaneous clicks on both OGSs. Accounting for the combinations of both true and extraneous clicks, we have
\begin{align}
\begin{split}
D_t = &(1 + p_\text{ap}^2) \left( \eta_t + \eta^A_t (1 - \eta^B_t) p_{\text{ec},t}^B + \eta^B_t (1 - \eta^A_t) p_{\text{ec},t}^A \right. \\
& \hspace{1.45cm}\left. + (1 - \eta_t) p_{\text{ec},t}^A p_{\text{ec},t}^B\right)
\end{split}
\end{align}
where we recall $\eta_t = \eta^A_t \eta^B_t$, $p_\text{ap}$ is the afterpulse probability, and $p_{\text{ec},t}^i$ the time-dependent extraneous count probability for link $i$ defined in Eq.~\eqref{eqn:ec_prob}. The block size is then readily determined by combining the total click $\sum_t f_s D_t / 2$, where the sum is over the visible time for a satellite overpass, and the factor $1/2$ accounts for sifting.

\begin{figure}[t!]
    \centering
    \includegraphics[width=1\columnwidth]{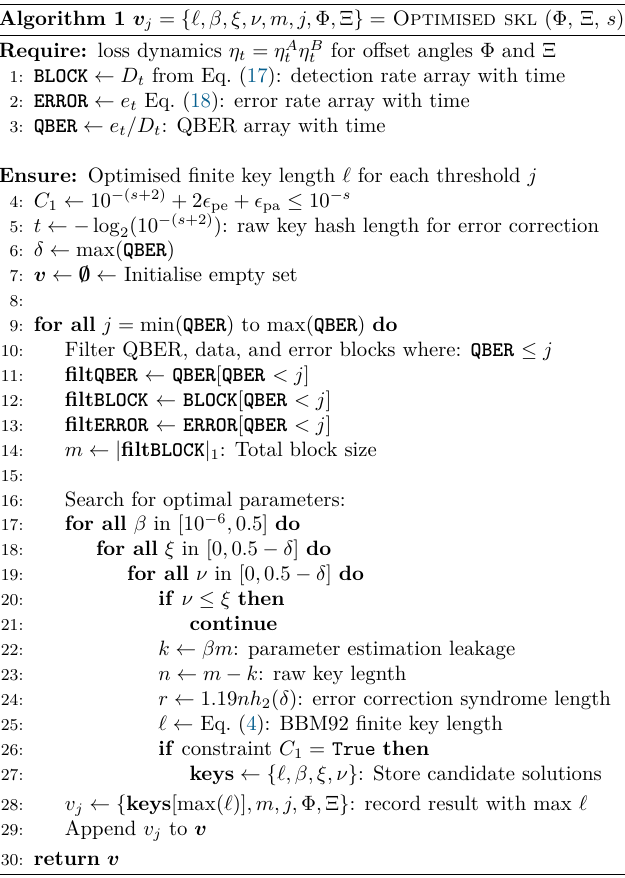}
    \caption{\textbf{Pseudocode for key rate maximisation}: The search grid is initialised by the nested for loops, which define suitable arrays for the optimisation parameters $\{\beta, \xi, \nu\}$ that by construction satisfy all the constraints. The internal loops are parallelised for computational speedup, with all optimisation parameters exported together with the key length, block size, QBER threshold, and trajectory angles. The notation $\vert\cdot\vert_1$ on line fourteen indicates the sum of the vector elements. The security level, $s$, in the work is set to six.}
    \label{fig:protocol_algo}
\end{figure}%

Similarly, to compute the QBER, we determine the error rate from the combination of extraneous and afterpulsing across both detectors, such that 
\begin{align}
\begin{split}
e_t = & \eta^A_t (1 - \eta^B_t) (1 - p_{\text{ec},t}^A) p_{\text{ec},t}^B \\
& + \eta^B_t (1 - \eta^A_t) (1 - p_{\text{ec},t}^B) p_{\text{ec},t}^A \\
& + (1-\eta_t) p_{\text{ec},t}^A p_{\text{ec},t}^B + p_\text{ap}^2 D_t/2 + \text{QBER}_\text{I} \eta_t \, ,
\label{eqn:qber}
\end{split}
\end{align}
where $\text{QBER}_\text{I}$ accounts for polarisation errors due to imperfection in the preparation of the Bell state. Typical values taken are 0.5-5\%. From this we calculate the number of photons, $E_t = f_se_t$, or the error block size. The time-dependent QBER is therefore determined from the ratio $E_t/D_t$. During the overpass, the QBER exhibits an inherent asymmetry due to differing probabilities of extraneous counts caused by different background light at each OGS site. This asymmetry also arises since each OGS-satellite link range is minimised at different times during the overpass. This effect is not an intrinsic property of dual-downlink systems, and is expected to largely cancel out when averaged over a distribution of satellite overpasses. 

With the data and QBER blocks, we have all the secure key metrics to determine the BBM92 secret key lengths. The full details of the key rate is provided in the main body (Section~\ref{subsec:protocol}), with full details on our key rate optimisation routine provided in the pseudocode in Fig.~\ref{fig:protocol_algo}.


\subsection{Thresholding models}
\label{subsec:thresholding}

\noindent
The finite key rate is heavily dependent on the threshold value $\delta$. To reason this, note that after partitioning the initial raw key string for parameter estimation and key generation, the threshold parameter $\delta$ serves as an upper bound on the estimated error in the finite keys. The protocol is designed to only proceed if the sample error rate is below a threshold $\delta$, preserving the security of the remaining data. However, a higher threshold leads to more bits sacrificed during error correction and privacy amplification. Therefore, a higher QBER threshold has the net effect of suppressing key generation. Conversely, a small threshold value would more often lead to aborting the protocol. The choice of $\delta$ is critical to both ensure security and maximise key generation, with its choice of value an influential factor in the finite key regime.

We consider two approaches for defining this threshold. First, $\delta=\text{max}(\texttt{QBER})$ defines a bound based on the maximum observed QBER over all time bins. This choice represents a conservative bound that guarantees protocol validity under the worst-case error conditions. By ensuring that all key generation blocks fall below this stringent upper limit, this model ensures robustness at the expense of efficiency, typically yielding shorter secure key lengths due to the larger fraction of bits consumed during error correction and privacy amplification. Second, 
\begin{align}
\delta = \frac{\sum_t E_t}{\sum_{t'} D_{t'}}
\end{align}
represents a weighted average of the block QBER. This formulation allows the threshold to vary dynamically with the selected block size, tailoring it to the observed statistical distribution of the channel. As a result, this second model can adopt smaller values for subsets of data corresponding to favourable link conditions where the losses are lower during the satellite overpass, which improves the overall key generation rate. This adaptive approach enables optimisation of the secret key length over block size and other operational parameters.

From a security perspective, both models are secure under the composable security framework. The accumulated key material is formed from independent temporal blocks whose individual error contributions are already attributed to a potential eavesdropper. Consequently, no additional information can be gleaned by an adversary, even under coherent attacks. Furthermore, since click statistics can be temporally scrambled, an eavesdropper cannot selectively target portions of the data corresponding to higher QBERs (low satellite elevation data). Thus, the choice between the two threshold definitions affects only the efficiency of key generation, not the fundamental security of the protocol.


%


\section*{Acknowledgements}

\noindent
We acknowledge support from the UK NQTP, the EPSRC Quantum Technology Hubs in Quantum Communications (EP/T001011/1) and Integrated Quantum Networks (EP/Z533208/1), the EPSRC International Network in Space Quantum Technologies (EP/W027011/1), and the Horizon Marie Sk{\l}odowska-Curie Fellowship (Project number: 101211639). We also acknowledge support from the UK Space Agency (NSTP3-FT-063, NSTP3-FT2-065, NSIP ROKS Payload Flight Model), Innovate UK (ReFQ project 78161, AirQKD project 45364, ViSatQT project 43037), EU QTSPACE (COST CA15220), and ESA Contracts 4000148330/25/NL/FGL/lf and 4000147561/25/NL/FGL/ss. We also acknowledge studentship support from Fraunhofer UK Research Ltd and the EPSRC Research Excellence Award (REA). \je{The Berlin team has been funded by the BMFTR (QR.X, QuSol, Hybrid++), Berlin Quantum, the Munich Quantum Valley, the Quantum Flagship (Millenion, PasQuans2), the DFG (SPP 2514), the Clusters of Excellence (ML4Q, MATH+), and the European Research Council (DebuQC).} The authors would like to thank Thomas Brougham for useful discussions and Alexander Lohrmann for feedback.

\clearpage

\onecolumngrid
\section*{Supplementary figures}

\renewcommand{\thefigure}{S\arabic{figure}}
\setcounter{figure}{0}

\subsection{Dual downlink channel model with $d=1000$~km}
\label{sec:dual_downlink}

\begin{figure*}[h!]
    \centering
    \includegraphics[width=0.86\linewidth]{FigS1.jpeg}
    \caption{\textbf{Dual-downlink channel model}. Loss and elevation as a function of time for a different satellite overpass geometries corresponding to offset distances $\Delta =$ 0~km (left column), 500~km (centre), and 1000~km (right). The black line illustrates the total system loss, the green dashed and dotted lines the elevation for link A and B respectively. The grey region marks time windows where the satellite's elevation for either link drops below $10^\circ$ and where no finite key is attainable. Note that the angle $\Xi$ dramatically reduces the available time to generate finite keys. All system parameters as detailed in Table~1 ($d=1000$~km).}
    \label{fig:grid}
\end{figure*}%

\subsection{Dual downlink channel model with $d=2000$~km}
\label{sec:dual_downlink}

\begin{figure*}[h!]
    \centering
    \includegraphics[width=0.86\linewidth]{FigS1b.jpeg}
    \caption{\textbf{Dual-downlink channel model}. Loss and elevation as a function of time for a different satellite overpass geometries corresponding to offset distances $\Delta =$ 0~km (left column), 500~km (centre), and 1000~km (right). The black line illustrates the total system loss, the green dashed and dotted lines the elevation for link A and B respectively. The grey region marks time windows where the satellite's elevation for either link drops below $10^\circ$ and where no finite key is attainable. Note that the angle $\Xi$ dramatically reduces the available time to generate finite keys. All system parameters as detailed in Table~1 (now with $d=2000$~km).}
    \label{fig:grid_2}
\end{figure*}%
%


\subsection{Optimisation parameters}
\label{supp:opt_params}

\begin{figure*}[h!]
    \centering
    \includegraphics[width=0.6\linewidth]{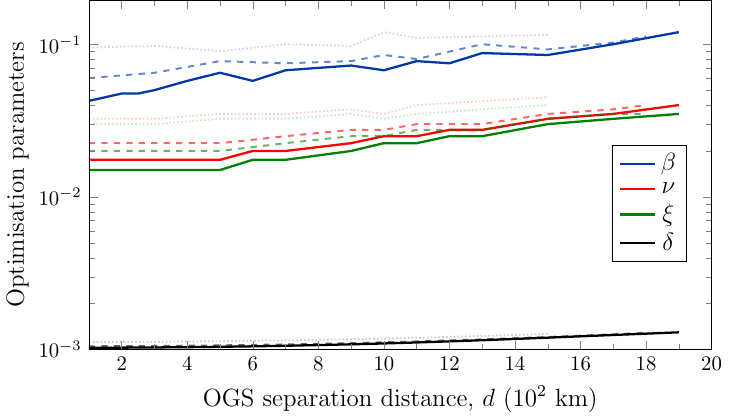}
    \caption{\textbf{Optimised finite-key parameters with OGS separation}. Variation of the four optimisation parameters $(\delta, \beta, \nu, \xi)$ with increasing separation between OGSs for a zenith-zenith ($\Phi=0$) satellite overpass with altitudes of 350~km (solid), 500~km (dashed), and 800~km (dotted). As separation, and hence the total channel loss, increases, the parameters adjust to maintain finite-key security: the tolerated error rate $\delta$ and sampling ratio $\beta$ rise to counter higher noise and statistical uncertainties, while the parameters $\nu$ and $\xi$ increase with increasing loss to maintain a tight bound on the excess error rate under low-count regimes. This illustrates the trade-off between key length and security that enables finite-key generation across varying link losses.}
    \label{fig:opt_param}
\end{figure*}%
%


\subsection{Overpass trajectory}
\label{supp:overpass_traj}

\begin{figure*}[h!]
    \centering
    \includegraphics[width=0.95\linewidth]{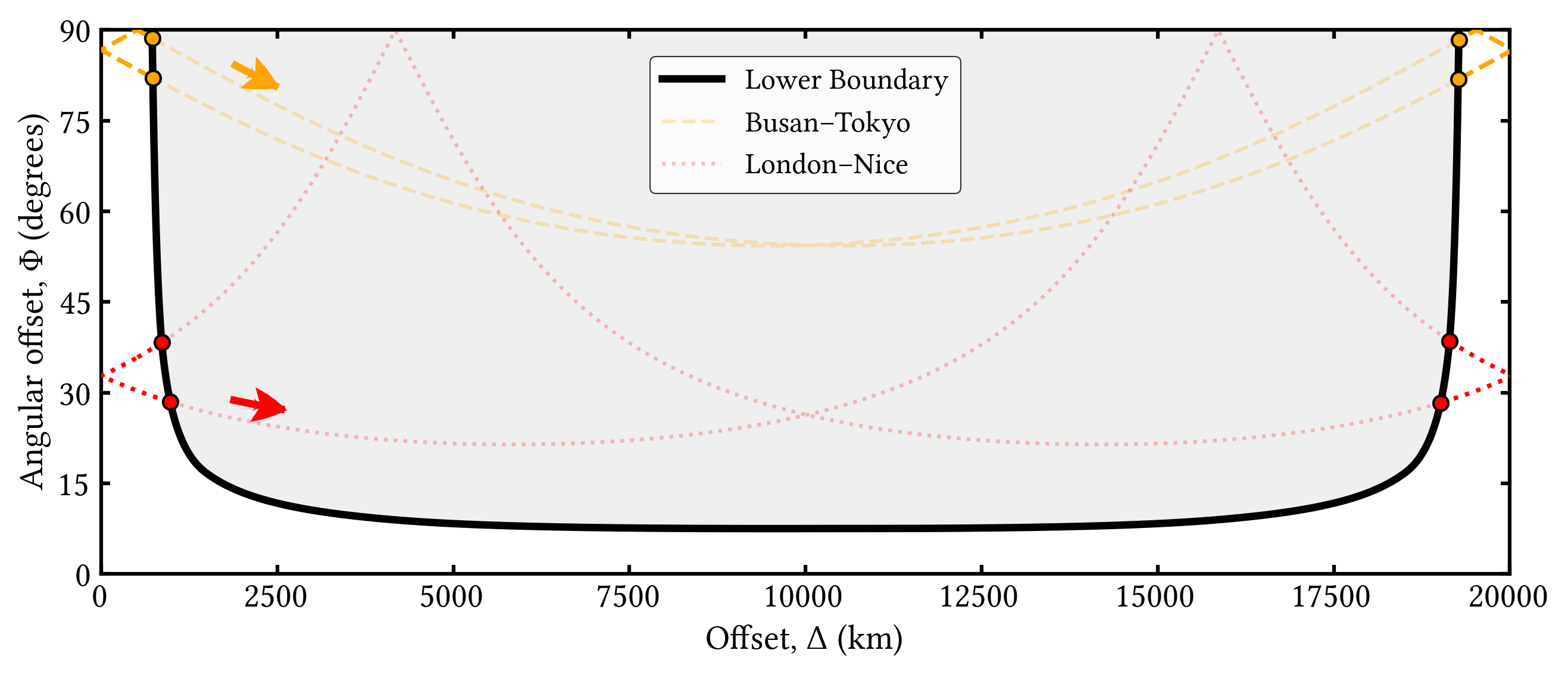}
    \caption{\textbf{Overpass trajectories in the ($\bm{\Delta, \Phi}$) parameter space for representative OGS pairings}. For a polar orbit, the satellite longitude $\gamma$ is uniformly sampled over a year, mapping the trajectory $(\Delta(\gamma),\Phi(\gamma))$ in the parameter space for each OGS pair. Shown here are trajectories for the London-Nice (green dotted) and Busan-Tokyo (cyan dashed) OGS pairs with angular offsets $\Phi^\text{LN}_{\Delta=0}=32.8^\circ$ and $\Phi^\text{BT}_{\Delta=0}=86.7^\circ$, respectively. The solid black illustrates the loss-limited boundary. Arrows indicate increasing $\gamma$, while markers denote intersections with the loss-limited boundary, beyond which secure key generation is precluded due to excessive channel loss. The London-Nice link predominantly samples small $\Phi$ (off-axis) geometries, whereas Busan-Tokyo accesses higher $\Phi$.}
    \label{fig:metropolitan_trajectories}
\end{figure*}%

\end{document}